\documentclass[a4paper,fleqn,usenatbib]{mnras}

\usepackage{mathptmx}
\usepackage[T1]{fontenc}
\usepackage{ae,aecompl}

\usepackage{graphicx}	% Including figure files
\usepackage{amsmath}	% Advanced maths commands
\usepackage{amssymb}	% Extra maths symbols

\title[Searching for Outflows in ULXs]{Searching for Outflows in Ultraluminous X-ray Sources Through High-Resolution X-ray Spectroscopy}

\author[P. Kosec et al.]{
P. Kosec$^{1}$\thanks{E-mail: pk394@cam.ac.uk},
C. Pinto$^{1}$,
A. C. Fabian$^{1}$
and D. J. Walton$^{1}$
\\
$^{1}$Institute of Astronomy, Madingley Road, CB3 0HA Cambridge, UK\\
}

\date{Accepted 2017 October 12. Received 2017 October 11; in original form 2017 August 29}

\pubyear{2017}

\begin{document}
\label{firstpage}
\pagerange{\pageref{firstpage}--\pageref{lastpage}}
\maketitle

\begin{abstract}

Ultraluminous X-ray sources are non-nuclear point sources exceeding the Eddington luminosity of a 10 Solar mass black hole. Modern consensus for a majority of the ULX population is that they are powered by stellar-mass black holes or neutron stars accreting well above the Eddington limit. Theoretical models of super-Eddington accretion predict existence of powerful outflows of moderately ionised gas at mildly relativistic velocities. So far, these winds have been found in 3 systems: NGC 1313 X-1, NGC 5408 X-1, NGC 55 ULX. In this work, we create a sample of all ULXs with usable archival high resolution X-ray data, with 10 sources in total, in which we aim to find more signatures of outflows. We perform Gaussian line scans to find any narrow spectral signatures, and physical wind model scans where possible. We tentatively identify an outflow in NGC 5204 X-1, blueshifted to 0.34c, which produces emission features with a total significance of at least 3$\sigma$. Next we compare ULXs with similar hardness ratios. Holmberg IX X-1 shows absorption features which could be associated with a photoionized outflowing absorber, similar to that seen in NGC 1313 X-1. The spectrum of Holmberg II X-1 possesses features similar to NGC 5408 X-1 and NGC 6946 X-1 shows O VIII rest-frame emission. All other sources from the sample also show tentative evidence of spectral features in their high resolution spectra. Further observations with the XMM-Newton and Chandra gratings will place stronger constraints. Future missions like XARM and Athena will be able to detect them at larger distances and increase our sample.

\end{abstract}


%%%%%%%%%%%%%%%%%%%%%%%%%%%%%%%%%%%%%%%%%%%%%%%%%%

%%%%%%%%%%%%%%%%% BODY OF PAPER %%%%%%%%%%%%%%%%%%

\section{Introduction}

Ultraluminous X-ray sources (hereafter ULXs) are a heterogeneous population of point-like non-nuclear X-ray sources with isotropic X-ray luminosities in excess of $10^{39}$ erg/s \citep[for a recent review of ULXs, see][]{Kaaret+17}. This means their luminosities exceed the Eddington luminosity of a stellar mass (10 $M_{\odot}$) black hole. ULXs tend to reside in star-forming regions and galaxies \citep{King+01}, and are often located in nebulae formed of mildly ionised gas, 10s to 100s pc large \citep{Pakull+06}.

The true nature of these compact object has been disputed for decades. Two explanations are most plausible. ULXs  could either be intermediate mass black holes ($\sim10^3-10^4 M_{\odot}$) accreting at sub-Eddington rates, or super-Eddington accretors of smaller mass. The first explanation might as well be the case for the more luminous objects at or above $\sim10^{41}$ erg/s \citep{Farrell+09,Webb+12}. However, modern consensus for a majority of the ULX population seems to prefer the second hypothesis \citep{King+09,Middleton+11,Sutton+13, Bachetti+13,Middleton+15}. In addition 3 of the ULXs were recently identified as neutron stars with extremely high accretion rates \citep{Bachetti+14,Furst+16,Israel+17}. At luminosities of several times $10^{40}$ erg/s, they surpass the Eddington luminosity by a factor of more than hundred (not taking possible geometric beaming into account).

Physical models of super-Eddington accretion (up to $\sim$100 times the critical mass accretion rate) imply a geometrically and optically thick disc around the central accretor \citep{Shakura+73,Poutanen+07,Fiacconi+17}. Theoretical simulations are also consistent with this concept \citep{Takeuchi+13,Narayan+17}. The main prediction of these models are massive radiatively driven outflows of ionised gas launched from regions close to the accretor, at mildly relativistic velocities. At high inclination angles from the rotation axis of the system, these outflows gradually become optically thick \citep[the opacity also depends on the mass accretion rate of the source, see][]{Middleton+15}. At a low angle from the rotation axis, the outflow is optically thin, effectively forming an evacuated funnel. For a current idea of structure of the system, see Fig. 13 of \citep{Pinto+17}. This means that the spectral hardness of a source could be directly related to the viewing angle under which it is being observed. At low inclination angles, the observer looks right into the evacuated funnel and sees the innermost parts of the thick accretion disc, which are thought to produce the hardest X-ray radiation. At higher angles, these regions are partially obscured by the disc itself and the outflowing wind. At very high inclination angles, only the Compton thick wind and outer parts of the accretion disc are visible, which produces a very soft X-ray spectrum - possibly an ultraluminous supersoft X-ray source (ULS) spectrum, and may resemble microquasars such as SS 433 \citep{Marshall+02} if it possesses jets. In general, it is thought that the higher the inclination angle, the softer the ULX appears \citep{Middleton+15,Feng+16,Pinto+17}.

ULX outflows are hence a major prediction of this theory, however they are much more difficult to observe in practice as it requires searching for weak absorption and emission features in X-ray spectra. This is particularly complicated by the fact that most ULXs reside in low metallicity environments which decreases the equivalent widths of X-ray lines. Additionally, even if winds are present in all ULXs, they might not be observable at all viewing angles. After the XMM-Newton and Chandra launch, strong residuals were spotted in ULX X-ray spectral fits  \citep[][and references therein]{Stobbart+06}. \citet{Middleton+15b} noticed these residuals appear to anticorrelate with spectral hardness, supporting the picture that they are associated with an outflow rather than reflection of a primary continuum. A confirmed detection was finally achieved in 2016 by \citet{Pinto+16} thanks to the use of high resolution spectroscopy with the Reflection Grating Spectrometer onboard XMM-Newton mission. As of August 2017, outflows have been discovered in 3 different ULXs, with the first two being NGC 1313 X-1 and NGC 5408 X-1, thanks to very high quality RGS data (300-700 ks each), both mildly relativistic at a speed of about 0.2 c. A weaker detection has been reported in NGC 55 ULX \citep{Pinto+17}. The most notable spectral features of these winds are neon X, iron XVII, oxygen VII and VIII emission lines. \citet{Walton+16} found evidence for a similar velocity outflow in iron K absorption.

In this work, we aim to search for similar signatures of outflows in a sample of ULXs using archival XMM-Newton data. This can be achieved by identifying any robust emission or absorption features in their X-ray spectra. We hope to collect as large a sample of ULXs with suitable (RGS) data and good enough statistics as possible (as of April 2017). We also try to span a range of different spectral categories of ULXs: soft ultraluminous, hard ultraluminous and broadened disc \citep{Sutton+13}.

The structure of the paper is as follows. In sections 2 and 3 we describe our data reduction techniques and methods for outflow detection, respectively. Section 4 contains the results of the in-depth search for emission and absorption features in ULX RGS spectra. Section 5 discusses the validity and implications of our results. We summarize the work done in section 6.

\section{Observations and Data Reduction}
\label{observations_reduction}

\begin{table}
	\centering
	\caption{Coordinates, distance and hardness ratio of ULXs used in this work. Column (1) lists the ULX name and its host galaxy, (2) and (3) contain its right ascension and declination, respectively and (4) its distance adopted in this work. Column (5) lists the average (absorption-corrected) hardness ratio for each source, which is defined as H/(H+S) where H is the X-ray flux in 2-10 keV band and S is the X-ray flux in 0.3-2 keV band.}
	\label{ulxdata}
	  \small\addtolength{\tabcolsep}{-2.6pt}
	\begin{tabular}{ccccc}
		\hline
		Object name&RA&Dec&Distance&Hardness\\
		 & & & & ratio\\
		 & hh mm ss& dd mm ss  & Mpc & \\
		(1)  & (2) & (3) & (4) & (5) \\
		\hline
		NGC 5643 X-1 & 14 32 42 & -44 09 36 & 16.9 & 0.639 \\		
		Holmberg IX X-1 & 09 57 53 & +69 03 48 & 3.7 & 0.603 \\
		NGC 4190 ULX1 & 12 13 45 & +36 37 55 & 2.9 & 0.569 \\
		M33 X-8 & 01 33 51 & +30 39 36 & 0.85 & 0.532 \\
		NGC 4631 ULX1 & 09 57 53 & +69 03 48 & 7.4 & 0.521 \\
		NGC 1313 X-2 & 03 18 22 & -66 36 04 & 4.3 & 0.509 \\
		IC 342 X-1 & 03 45 56 & +68 04 55 & 3.3 & 0.500 \\
		NGC 5204 X-1 & 13 29 39 & +58 25 06 & 5.6 & 0.403 \\
		Holmberg II X-1 & 08 19 29 & +70 42 19 & 3.3 & 0.242 \\
		NGC 6946 X-1 & 20 35 01 & +60 11 31 & 5.6 & 0.237 \\
		\hline
	\end{tabular}
\end{table}

All observations in this work were carried out by the XMM-Newton \citep{Jansen+01} satellite. We use data from both the European Photon Imaging Camera (EPIC) PN \citep{Struder+01} and the Reflection Grating Spectrometer (RGS) detectors \citep{denHerder+01}. We select all objects with suitable and public RGS data that do not yet have a reported detection of an outflow. Suitable RGS data means the object is correctly aligned with an appropriate roll angle and not offset by more than 2 arcmin from the optical axis of the spectrometer. The source also needs to have enough counts (at least $\sim$1000 in total) and a high enough count rate (above $\sim$0.1 count/s with PN in the 0.3-10 keV band) to be able to detect any spectral features. In the end, our sample consists of 8 ULXs with different spectral properties: Holmberg IX X-1, Holmberg II X-1, NGC 1313 X-2, NGC 4190 ULX1, NGC 5204 X-1, NGC 5643 X-1, NGC 6946 X-1 and M33 X-8. We also study NGC 4631 ULX1 but given its brightness, detection of any spectral features does not seem feasible. IC 342 X-1 is bright enough for the analysis but its RGS data are contaminated by another X-ray source in the source region.

\begin{table*}
	\centering
	\caption{Log of the observations used in this work. Column (1) lists the ULX name and its host galaxy. Column (2) contains the number of exposures of the ULX by XMM-Newton and (3) the total clean RGS exposure (per detector) of all pointings used in this analysis, after subtracting solar flaring periods. Column (4) lists all exposures of sources used in this paper. }
	\label{xmmobsdata}
	\begin{tabular}{cccc} % four columns, alignment for each
		\hline
		Object Name&Number of exposures&Total duration&Observations used\\
		 & & ks & Obs ID \\
		(1)  & (2) & (3) & (4) \\
		\hline
		NGC 5643 X-1 &  1 & 108 & 0744050101 \\
		Holmberg IX X-1 & 9 & 159 & 0112521001 0112521101 0200980101 0693850801 0693850901 \\
		 & & & 0693851001 0693851101 0693851701 0693851801 \\
		NGC 4190 ULX1 & 3 & 43.3 & 0654650101 0654650201 0654650301 \\
		M33 X-8 &       2 & 20.5 & 0102640101 0141980801 \\
		NGC 4631 ULX1 & 1 & 44 & 0110900201 \\
		NGC 1313 X-2 &  2 & 102.5 & 0764770101 0764770401 \\
		IC 342 X-1 &    2 & 87.5 & 0693850601 0693851301 \\
		NGC 5204 X-1 &  9 & 162.1 & 0142770101 0142770301 0150650301 0405690101 0405690201 \\
		 & & & 0405690501 0693850701 0693851401 0741960101 \\
		Holmberg II X-1 & 2 & 99.5 & 0200470101 0561580401 \\
		NGC 6946 X-1 &  1 & 109.9 & 0691570101 \\
		\hline
	\end{tabular}
\end{table*}

All objects studied in this analysis are listed with their properties such as the distance and coordinates in Table \ref{ulxdata}. We obtain the source distances by averaging newer measurements shown in the NED database. It should be kept in mind that due to low distances of ULXs, the assumption of exact distances does not actually affect the following analysis. Absolute luminosity measurements would be affected, but they are not used in this work. The coordinates and source names are obtained from SIMBAD. We calculate an average hardness ratio for each source. We fit the broadband PN spectrum with the standard continuum model (as described in Sect. \ref{continuum_spectral_fitting}) and calculate the X-ray flux between 0.3-2 keV (S) and 2-10 keV (H). The hardness ratio is then defined as H/(H+S). Finally we use all individual observations of a source to calculate an average hardness ratio. This way the ratio takes into account different obscuration levels for each source.

The observational info such as exposures and Obs IDs are shown in Table \ref{xmmobsdata}. For some of the objects in our sample, more XMM-Newton data are available but they are unsuitable for RGS analysis due to low S/N ratio or bad roll angle.

All the data were downloaded from the XSA archive and reduced with a standard pipeline using SAS v15.0, CalDB as of April 2017. High background rate periods were filtered out, with filtering threshold of 0.5 counts/sec for PN detectors and 0.25 counts/sec for RGS spectrometers. Events of PATTERN <= 4 (single/double) were accepted for PN data. The source regions for PN were selected as circles centred on the ULX with a radius of 30 arcsec, to avoid chip gaps. The background regions were circles located in the same region of the chip, avoiding the copper ring, chip gaps and out of time events from the source. We used the default selection of source and background regions in RGS where possible. In several cases we had to do custom background selection for the RGS data through RGS masks, for example in NGC 5643 X-1 due to contamination of default background regions by other sources.
Reduced data were converted from the OGIP to SPEX format using the Trafo\footnote{http://var.sron.nl/SPEX-doc/manual/manualse100.html} tool. PN data were grouped by a minimum of 25 counts per bin using the GRPPHA tool (to achieve Gaussian statistics) and RGS data were binned by a factor of 3 directly in SPEX (to oversample the instrument resolution by a factor of about 3).
All RGS observations were later stacked for plotting purposes, making sure the selection regions were identical to avoid any energy shifts. The PN observations were also stacked using the epicspeccombine task within SAS.
The spectral range used was 0.3 to 10 keV for EPIC PN (limited by the effective area and calibration uncertainties) and 7 to 26 \AA\ (or 20 \AA\ where necessary) for RGS data (limited by the background).

\section{Methods}

\subsection{Continuum spectral fitting}
\label{continuum_spectral_fitting}

We use SPEX \citep{Kaastra+96} for spectral fitting, and Cash statistics \citep{Cash+76} as there are not enough data points per bin in RGS data for a $\chi^{2}$ analysis. All model parameters are checked extensively with a proper error search in case there are multiple minima in the C-stat function.

We fit PN data between 0.3 and 10 keV with a \textsc{hot*(pow+mbb+bb)} model where possible. \textsc{hot} reproduces mostly neutral (gas temperature of about 0.5 eV) galactic ISM absorption plus any additional absorption near the source itself \citep[see][for an example of similar usage of the model]{Kaastra+06}. \textsc{pow} (standard powerlaw) and \textsc{mbb} (colour corrected black body) model X-ray emission from close to the compact object. \textsc{bb} is a standard blackbody model which represents the photosphere at larger distances from the accretor. Similar X-ray continuum models are used by \citet{Gladstone+09} and \citet{Walton+14}.

The RGS continuum is easier to fit with case-by-case approach. Where counts are sufficient (long observations of Holmberg II X-1 and IX X-1, and stacked data of NGC 5204 X-1), we fit RGS data with the standard model, only checking PN data for any discrepancies. For sources with less counts, we take the model from PN spectral fitting, freeze all its parameters except for the overall normalization (using parameter coupling) and fit this model to RGS data of the source. The only exception from these 2 approaches is NGC 5643 X-1, where a simple \textsc{hot*pow} spectral model provided a satisfactory fit.

\subsection{Line search}

We perform a Gaussian line search in SPEX to find any spectral lines, both absorption and emission, that a possible outflow might have imprinted on the X-ray continuum of the source. We use the continuum spectral model obtained following the procedure above, and add another component which is a single Gaussian line at a certain energy, with a predefined spectral width. The original continuum is kept frozen except in the case of high quality datasets, where we are able to free its overall normalisation without breaking the fit in SPEX. Then we fit the normalisation of the added line. The normalisation can be both positive or negative, to reproduce an emission and absorption line, respectively. The width of the line is calculated based on a grid of velocity dispersions of the gas that we want to describe. 

After the fit, $1\sigma$ error on the normalisation of the line is calculated and saved as well as the $\Delta$C-stat improvement of the fit compared to original C-stat value. The approximate significance of the line in $\sigma$ can be calculated as a ratio of the normalization to the average of its $1\sigma$ errors (upper and lower). It should however be kept in mind that the $\Delta$C-stat value is the main guideline as to whether the detection is significant or not along with dedicated Monte Carlo simulations (see Section \ref{MCSims})

We proceed as described above for a grid of line energies spanning the whole energy range. We adopt about 2000 energy steps for the energy band which makes it computationally reasonable and samples the RGS resolution well enough. Typically, we search for lines with 3 different velocity dispersions (line widths) simulating different physical scenarios: 500 km/s, 1000 km/s and 5000 km/s. In the end we obtain a table of $\Delta$C-stat values, normalisations and significances of Gaussian lines for the whole energy band.

\subsection{Line significance through MC simulations}
\label{MCSims}

It should be kept in mind that the $\Delta$C-stat values give only approximate $\sigma$ significances because they do not take into account the "look-elsewhere" effect. Only simulations can give a rigorous answer about the actual significance of a line. In this case a fake spectrum is simulated with the same response matrix and using the template model of the continuum as the original spectrum. Then an identical line search is performed as for the original data. The simulation is repeated as many times as desired, and we count the number of occurrences of lines in simulated data at the same or higher significance as the $\Delta$C-stat of the line found in real data. The real significance (probability) of a line found in the original data is the ratio of simulations that do not find a line with the same or higher $\Delta$C-stat value to all performed simulations. For a 3$\sigma$ search (99.7 \% probability), the need arises to perform well over 1000 simulations. Computational time necessary for the task can be of order of 1000 hours, so a rigorous analysis of all objects in the sample using this method is not currently feasible and we perform it only on the most promising candidates.

\subsection{Spectral model search}
\label{spectral_model_search}

In the last part of this work, we use a physical model of ionised plasma to describe a potential outflow. The first model we make use of is \textsc{xabs} (in SPEX), which reproduces absorption by photoionized gas. This is mainly motivated by detection of an absorption line around 12.5 \AA\ in Holmberg IX X-1 and around 15 \AA\ in Holmberg II X-1, a similar spectral feature to what was found in NGC 1313 X-1 and NGC 5408 X-1 \citep{Pinto+16}, albeit at smaller $\Delta\lambda$. This could imply a smaller (or more projected) outflow velocity.

The method is similar as the Gaussian line search - a \textsc{xabs} component was created in addition to the continuum fit, and its blueshift, the velocity of the \textsc{xabs} component with respect to us, was varied in a grid between 0 and 0.3c. At every point of the grid, the model was fitted while keeping the continuum fit frozen except for its overall normalisation. The column density and the ionisation parameter of the absorbing gas was kept free. The typical velocity dispersion of gas within the outflow was kept frozen and was varied between computing runs to describe different gas outflow properties. We adopt solar metallicity of elements while using the model.

To fit any potential emission features, we use a collisionally ionised emission model, \textsc{cie} in SPEX. The model is similar to \textsc{mekal} in XSPEC, but with many recent updates to atomic data \citep[particularly after the Hitomi observation of the Perseus cluster, see][]{Hitomi+16}. This model could reproduce plasma emission from a shocked region in case the ULX possesses jets, just as observed in SS433. We add a \textsc{cie} component to the continuum model and fit its temperature and normalization for a specific blueshift. To apply blueshift to the \textsc{cie} component, we have to make use of the \textsc{reds} model in SPEX. We also freeze the dispersion velocity of the collisionally ionised gas to a particular value, e.g. 250 or 1000 km/s. Subsequently, we vary the blueshift of the \textsc{cie} component in a grid between 0c and 0.4c. We find any potential statistical fit improvements by checking the $\Delta$C-stat value at each step. Then we can obtain the real significance of a detection with Monte Carlo simulations (similar to Gaussian line search MC simulations).

\section{Results}

\begin{figure*}
	\includegraphics[width=17.5cm]{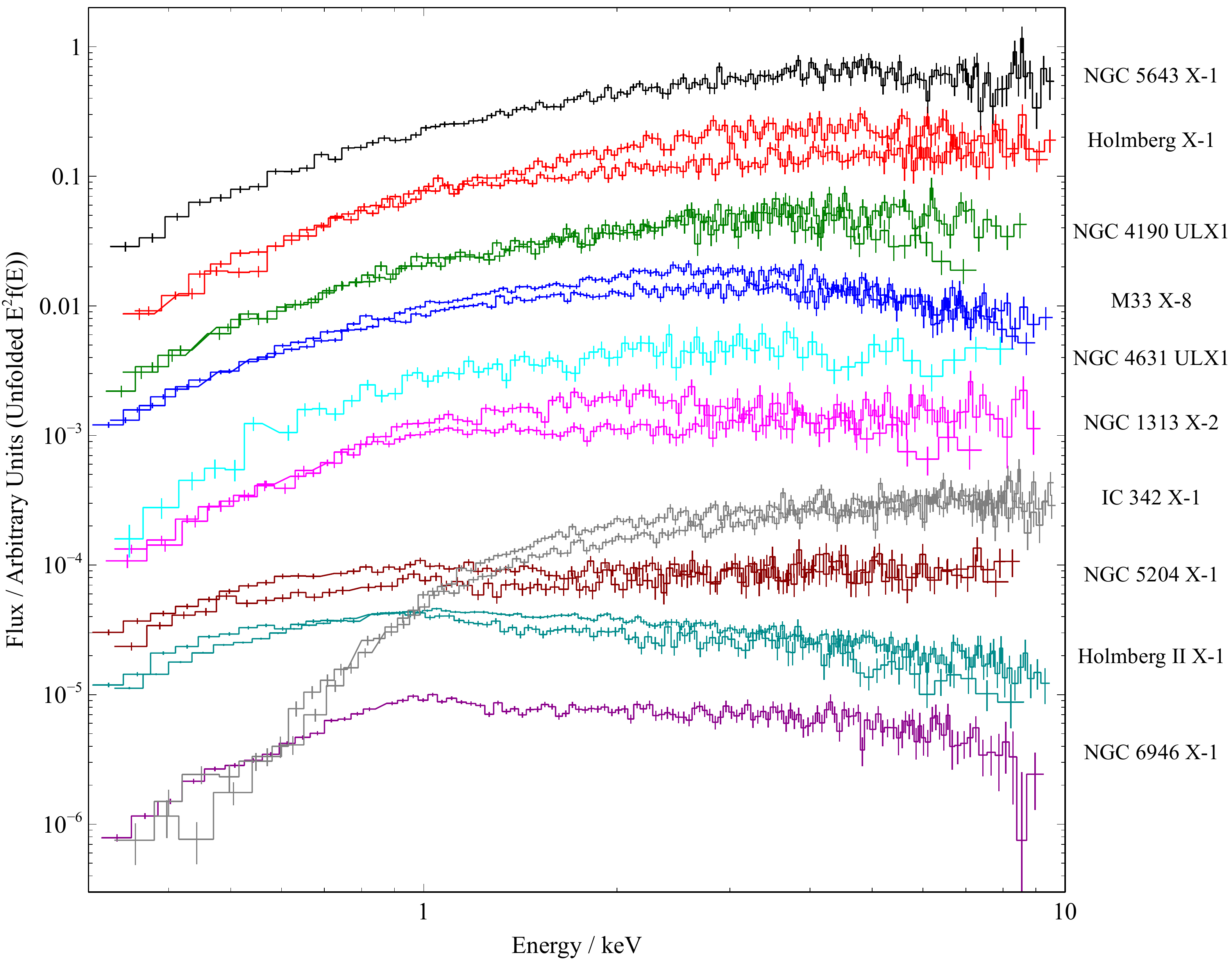}
    \caption{Unfolded (E$^{2}$f(E)) PN spectra of ULXs used in this analysis, between 0.3 and 10.0 keV. Y axis is in arbitrary units (fluxes are rescaled for plotting purposes). Sources are ordered by increasing hardness H/(H+S), where H is the 2.0-10.0 keV flux and S is the 0.3-2.0 keV flux, from bottom to top of the plot (see also Sect. \ref{observations_reduction}). Different objects are plotted in different colours, and where multiple observations are available, we plot only the 2 observations with most different hardness ratios, both with the same colour. Sources plotted are: black - NGC 5643 X-1, red - Holmberg IX X-1, dark green - NGC 4190 ULX1, dark blue - M33 X-8, cyan - NGC 4631 ULX1, pink - NGC 1313 X-2, grey - IC 342 X-1, dark red - NGC 5204 X-1, dark cyan - Holmberg II X-1, purple - NGC 6946 X-1. The spectra are not corrected for absorption.}
    \label{fig:PN_all}
\end{figure*}

All the broadband PN spectra of the sources from our sample are shown in Fig. \ref{fig:PN_all}. The detailed results from the RGS Gaussian line search for each source are shown in Fig. \ref{RGS_all} in Appendix \ref{RGS_plot}. Detailed comparisons are done for sources with similar hardness ratio such as Holmberg IX X-1 versus NGC 1313 X-1, and Holmberg II X-1 versus NGC 5204 X-1 versus NGC 5408 X-1. In this section we discuss the analysis performed for each source individually.

\subsection{Holmberg IX X-1}

There are 9 observations of this well known ULX in total, including a very long exposure (0200980101) at 120 ks. Unfortunately, the source was caught in a lower flux state in this observation so the total count number is not as good as expected, but still much better than any other observation. Initially, we fit the broadband PN spectrum for each observation separately. The spectral shape can usually be fitted with a \textsc{hot*(bb+mbb+pow)} model (see Section \ref{continuum_spectral_fitting} for more information). We start by performing a full line search on all observations separately, however we find that only the 120 ks observation has good enough S/N ratio to be analysed separately.

We first analyse the highest quality observation (0200980101). Initially we fit the broadband continuum of the source between 0.3 and 10 keV using PN data. We use a double blackbody (a simple blackbody model plus a color-corrected blackbody) plus a powerlaw model. All these components are absorbed by neutral absorption, which is reproduced by the \textsc{hot} component in SPEX. We obtain the following results: The temperatures of the blackbody components are $T_{BB} = 0.204 \pm 0.10$ keV and $T_{MBB} = 3.94_{-0.15}^{+0.16}$ keV. The powerlaw slope gamma is $2.0 \pm 0.2$ and the absorber column density near the source plus galactic ISM absorption is $1.36_{-0.16}^{+0.17}*10^{21}$ cm$^{-2}$. 

\begin{figure*}
	\includegraphics[width=18cm]{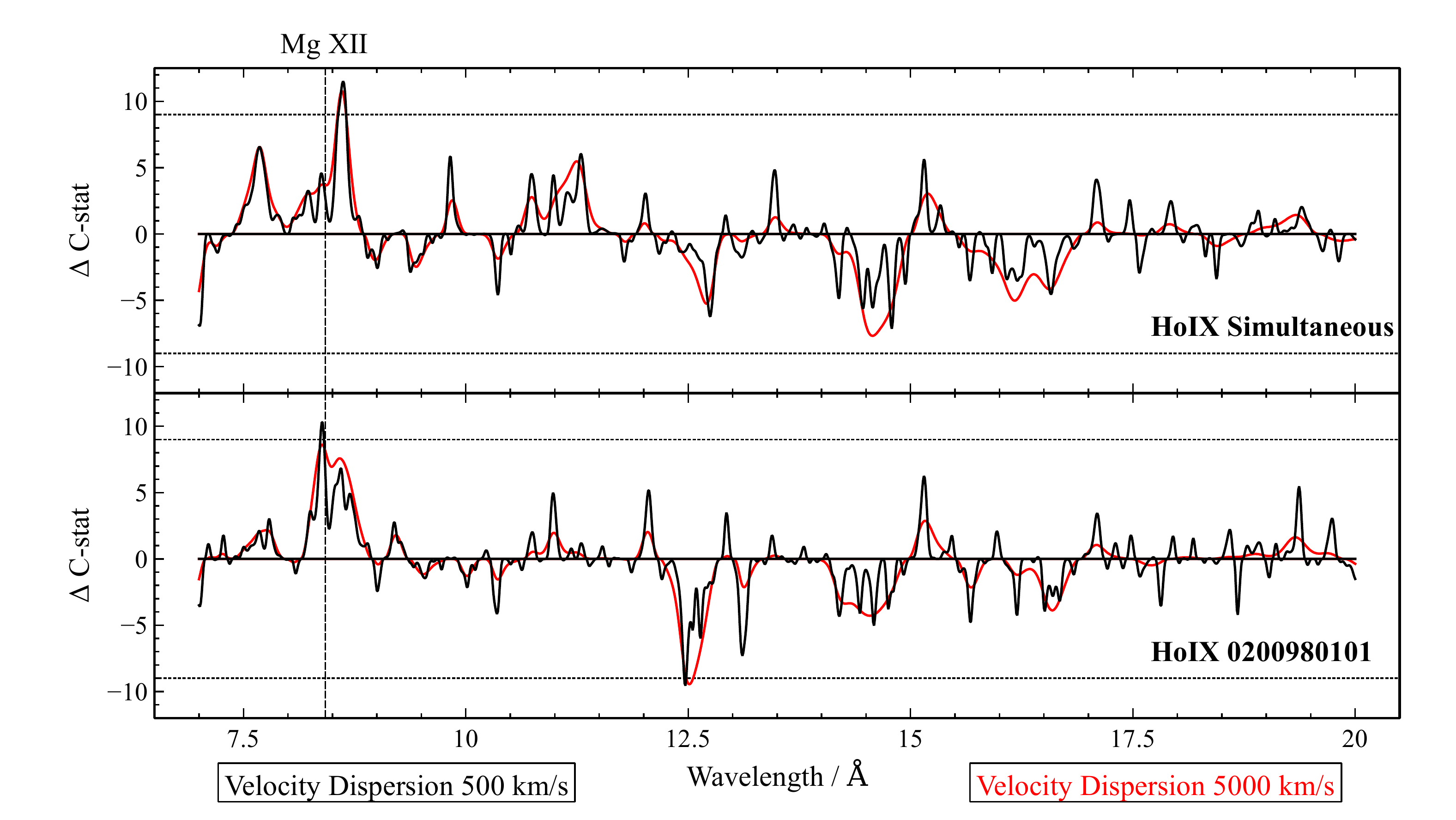}
    \caption{Gaussian line search results for Holmberg IX X-1. The Y axis is defined as $\Delta$C-stat times the sign of normalisation of the line to show the difference between absorption and emission features. Horizontal dotted lines show the values of $\Delta$C-stat=9 and -9. }
    \label{HoIX}
\end{figure*}

The RGS spectrum has about 7000 source counts in total, but the background is quite high and limits our analysis to the 7.0 to 20.0 \AA\ region only. The line search does not find any lines at very high significance, see Fig. \ref{HoIX}. There is an emission spectral feature with strength of over 10 $\Delta$C-stat at 8.5 \AA\ and an absorption feature with $\Delta$C-stat of over 9 located at 12.5 \AA. Otherwise the spectrum is clean of any strong features.

The other 8 observations each individually do not have enough counts to do a full line search therefore we did a combined line search using all 9 observations. The source varies between observations, hence we cannot use a single broadband spectral model. We group the observations into several spectral groups based on their flux and the time of observation. These groups share the same spectral model. We end up with 6 different spectral models: group 1 - observation 0112521001 + 0112521101, group 2 - 0200980101, group 3 - 0693850801, group 4 - 0693850901 + 0693851001, group 5 - 0693851101 and group 6 - 0693851701 + 0693851801. We fit the broadband spectral model for each group separately based on their PN spectra. We then follow by a simultaneous Gaussian line search on all these fits without any stacking of the data itself. This way we are looking for spectral features that are present in all of the observations, taking into account the variability of the source itself.

The simultaneous line search does not bring any conclusive results either. The 12.5 \AA\ absorption feature gets weaker than in the 0200980101 observation (now at $\Delta$C-stat=6), but does not disappear completely hence it could still be present in some observations (and not present in others). On the other hand, the emission feature at 8.5 \AA\ is now stronger at $\Delta$C-stat of almost 12, therefore it must be present at least in a fraction of other observations. Its width also diminishes. Finally a not very strong, but quite wide absorption feature appears at around 14.5 \AA\ with $\Delta$C-stat of about 8.

\subsection{Holmberg II X-1}

There are 2 observations of Holmberg II X-1 usable for RGS analysis, one high quality pointing with 56 ks of clean exposure and the source in a higher flux state (3 count/s in PN), and a second one with 44 ks and the source in a lower state (1.2 count/s).

First we analyse the long observation. The data quality is very good with almost 13000 source counts in RGS and we are able to fit the continuum based just on RGS data (hence a PN continuum is not necessary). A reasonable fit is obtained with a single blackbody plus powerlaw model with a temperature of $0.25_{-0.03}^{+0.04}$ keV and a powerlaw slope of $1.7 \pm 0.3$. We perform a Gaussian line search over the 7 to 26 \AA\ band where the continuum flux is significantly above the background. The data quality allows us to unfreeze the overall normalisation of the continuum and we fit it in addition to Gaussian line normalisation. This could increase the line strength found by the line search, but potentially can break the fitting program unless the data quality is high enough.

\begin{figure*}
	\includegraphics[width=18cm]{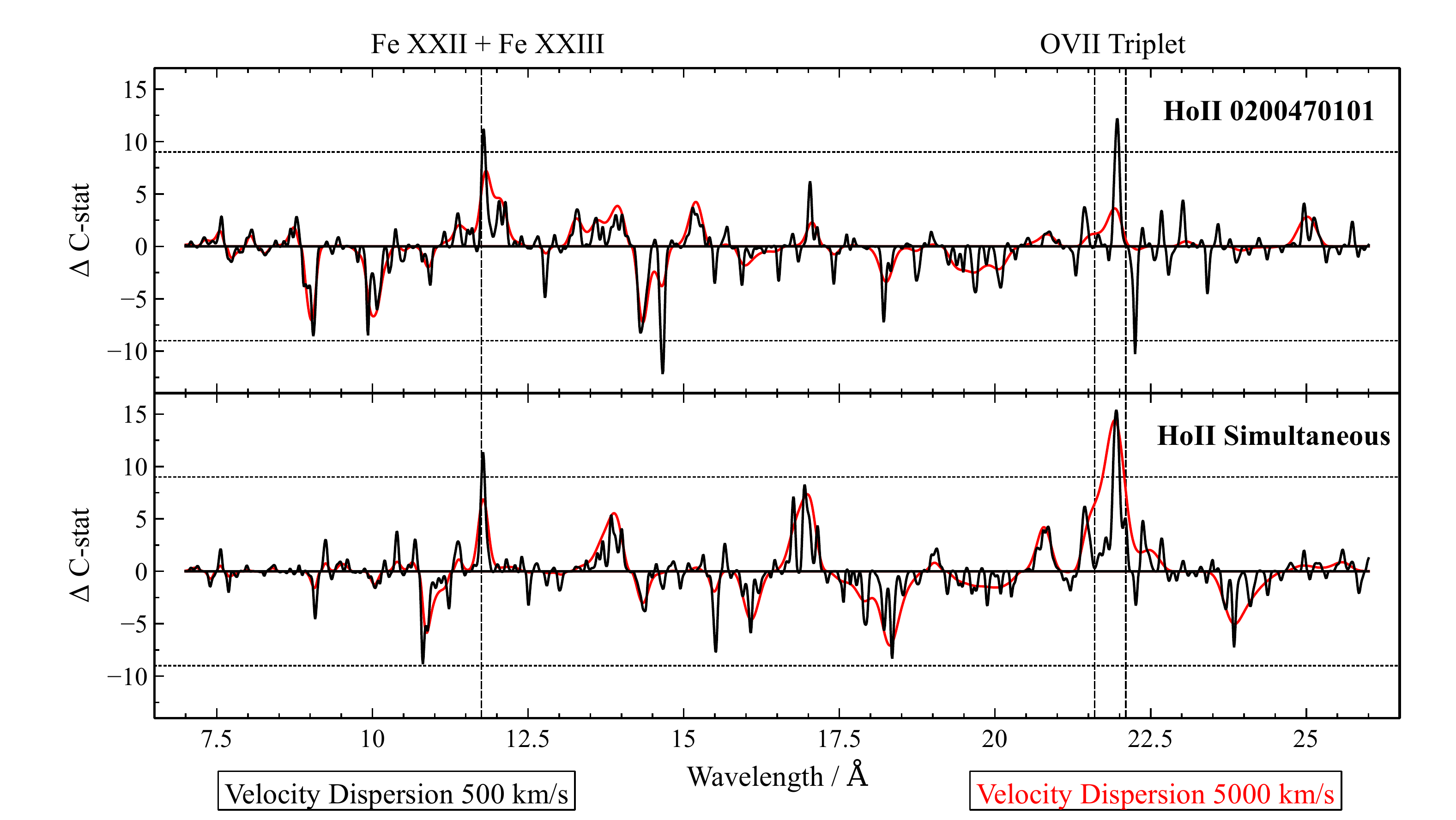}
    \caption{Line search results for Holmberg II X-1. Axes are defined as in Fig. \ref{HoIX}.}
    \label{HoII}
\end{figure*}

The analysis shows multiple strong residuals, see Fig. \ref{HoII}. Two emission features are found at 11.8 \AA\ with $\Delta$C-stat=11, and at 22 \AA\ with $\Delta$C-stat=12. There are also multiple absorption features: an absorption doublet around 14.5 \AA\ and $\Delta$C-stat of 12 and 8, around 22.3 \AA\ with $\Delta$C-stat=10, and two weaker but broad features at 9 and 10 \AA\ with $\Delta$C-stat of 8.

To quantify the actual probability of the detected spectral features we perform 1000 simulations of fake spectra as described in Section \ref{MCSims}. Out of the total number of simulations, 2 have 2 absorption lines with $\Delta$C-stat higher than the real data detections. This suggests the combined significance of our detection is about $3\sigma$.

The second observation is much shorter and the source is at lower flux, hence it is not good enough for an individual RGS analysis, particularly for absorption line search. We therefore perform a simultaneous line search in both observations at once. As usual, both spectra have their continua fitted separately, to which we add the same Gaussian line. The line search finds that the absorption features at 14.5 \AA\ weaken considerably, suggesting they are not present in the second spectrum. Alternatively they could be dominated by the background. However, some of the emission features remain. The 11.8 \AA\ line stays at the same $\Delta$C-stat=11 level meaning it is present in the second observation at least partially, and the 22 \AA\ line increases in significance to $\Delta$C-stat of over 15 suggesting it is definitely present in both spectra.

\subsection{NGC 5643 X-1}

NGC 5643 X-1 is the hardest ULX in our sample and has a single long exposure which is well centred for RGS use. It is also the most distant ULX studied here at 17 Mpc, so its count rate is relatively low ($\sim$0.25 counts/s in PN data). The broadband 0.3 to 10 keV spectrum can be well fitted with a blackbody and a second colour-corrected blackbody at temperatures of $0.31 \pm 0.03$  and $2.29 \pm 0.07$ keV, respectively (no powerlaw needed).

However, fitting the RGS data with this model results in a relatively poor fit and is not re-normalisable as there is slope difference between the PN and RGS data, most likely caused by high RGS background. We therefore adopt a completely different, simple powerlaw model for our line search analysis. This results in a reasonable fit (C-stat of 514 for 432 degrees of freedom), with a hard powerlaw coefficient of $1.28 \pm 0.14$. Due to strong background above 20 \AA\ in RGS data, we are forced to perform the line search in 7 to 20 \AA\ range. 

The line search finds a potential emission line at $\sim$13.8 \AA\ at $\Delta$C-stat of almost 12. There are more residuals between 18 and 20 \AA\ but these are most likely caused by the background or random fluctuations. The line search results are shown in Fig. \ref{RGS_all}.

\subsection{NGC 4190 ULX1}

There are 3 observations of the source in total, but given their statistics an individual line search would not likely be successful. Initially we fit the PN data with our standard model. One of the PN spectra is not usable as the detector has been swamped with flares for practically the whole exposure time, but we fit the other two observations with a double blackbody (first observation) and a blackbody plus powerlaw (second observation) model. For the observation without any continuum model, we use the PN model from the first observation which was taken only 2 days apart and their RGS continua are super-imposable. 

Now we renormalise the PN continua to the RGS spectrum and search for spectral features in all 3 observations simultaneously. We have about 2500 RGS counts in total. Overall the search does not find any very significant features, but we find a broad emission residual peaking at 18.7 \AA\ with $\Delta$C-stat=12. The feature is suspiciously broad, but at the same time the counts are well above the background level in this spectral range. There are other residuals found by the line search but none are stronger than $\Delta$C-stat=10. The line search results for NGC 4190 ULX1 can be found in Fig. \ref{RGS_all}.

\subsection{M33 X-8}

M33 X-8 is atypical for our sample as it only barely reaches the luminosity to be considered a ULX. We possess 2 observations of the source, which are short but the source is very bright thanks to its proximity (count rate of about 5 count/s with PN). We fit the broadband spectrum as usual and find a good fit with the standard double blackbody and a powerlaw model.

The RGS spectra have about 7000 counts combined. We renormalise the PN spectra to the RGS level and search for any spectral features simultaneously in both observations between 7 and 26 \AA. The simultaneous search finds an emission feature at 12.5 \AA\ with a strength of about $\Delta$C-stat=11, but other than that the spectrum is relatively clean. It is very unlikely that this feature is caused by the background given the high count rate of the source. The feature found by our line search is shown in Fig. \ref{RGS_all}.

\subsection{NGC 1313 X-2}

We are using the newly public data on NGC 1313 X-2, which is the only existing data of this source well centred for an RGS analysis with sufficient exposure time. Two exposures of 110 ks in total show that the source has varied significantly - increasing the PN count rate from 0.24 count/s during the first, longer observation, to 0.41 count/s during the second observation. We fit the first spectrum with a full broadband model, while the second one only requires a double blackbody (\textsc{bb} + \textsc{mbb}) fit.

Each observation only contains about 800 RGS counts (both RGS detectors combined). The background is stronger than the source above $\sim$20 \AA, so we search in the usual 7 to 20 \AA\ range. We fit the RGS spectra and renormalise them to avoid any constant residuals in the line search. Then we perform a simultaneous line search for both observations at once, i.e. we are looking for residuals that are present at both times. We find absorption residuals at 8, 9.6 and 13 \AA, and a bit weaker but a very broad feature at 15.5 \AA. There are potential emission line features at 10, 14.5 and 16.2 \AA, and it seems that the background affects our results at least above 18 \AA. A plot of these features can be seen in \ref{RGS_all}.

\subsection{NGC 5204 X-1}

\begin{figure*}
	\includegraphics[width=18cm]{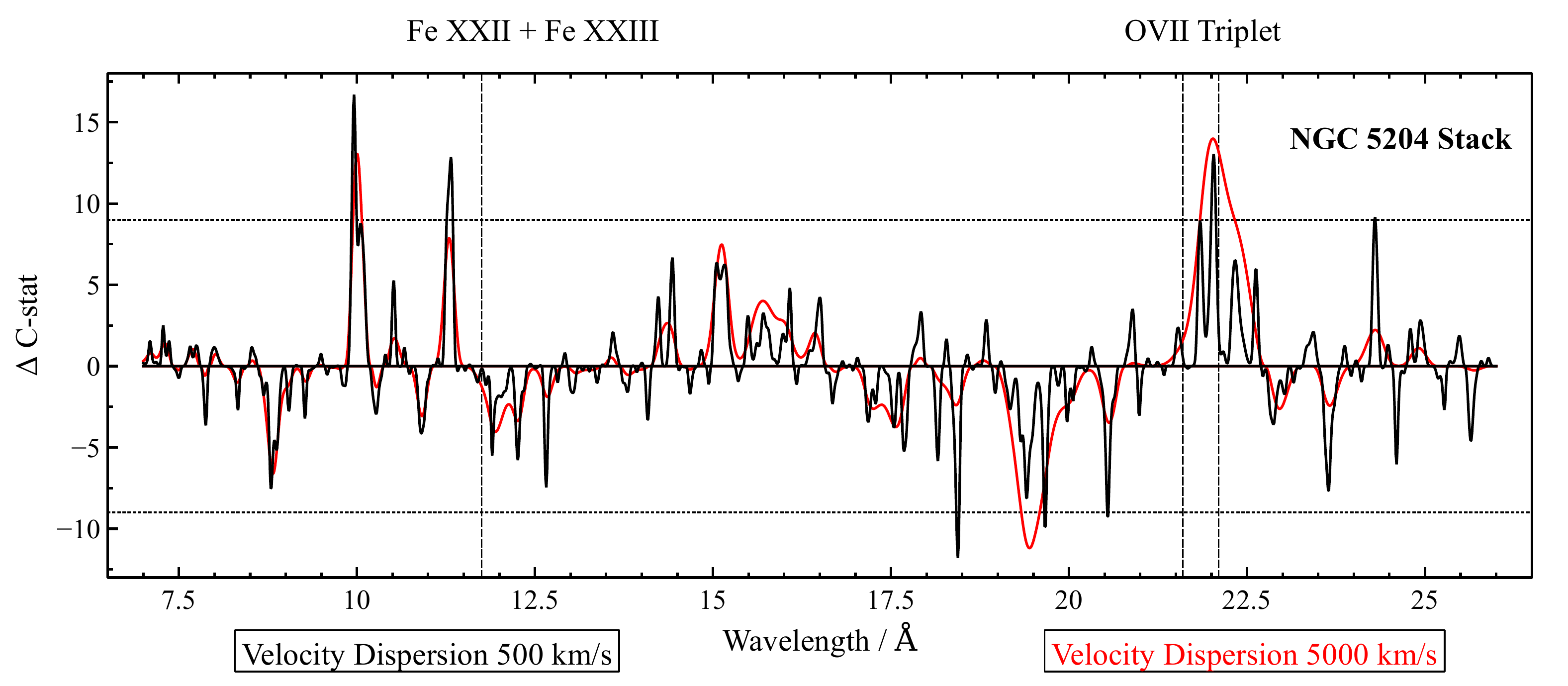}
    \caption{Line search results for NGC 5204 X-1. Axes are defined as in Fig. \ref{HoIX}.}
    \label{N5204}
\end{figure*}

There are 9 observations of NGC 5204 X-1 in total that are pointed well enough for RGS analysis, with a total raw exposure of 250 ks (160 ks filtered). The object varies in flux between about 0.5 count/s to 1.2 count/s (with PN), but as the observations are rather short, none of them is of high enough quality to be searched individually. We perform a custom RGS data reduction and extraction for each observation to avoid any possible errors such as energy shift and then stack them into a single spectrum. This simplifies the analysis greatly. We are able to get away with stacking despite the long-term variability of the source. It varies in normalisation but its spectral hardness does not change considerably. The spectrum is fitted with a single blackbody plus a powerlaw model (the second blackbody not necessary) with a temperature of $0.21^{+0.06}_{-0.05}$ and a gamma index of $1.9^{+0.4}_{-0.3}$.

We have 9000 counts in total which gives very good statistics compared to some other sources in our sample. The line search (see Fig. \ref{N5204}) finds a very strong emission feature at 10 \AA, with $\Delta$C-stat of over 16, and weaker features at 11.3 \AA\ ($\Delta$C-stat=13) and at 22 \AA ($\Delta$C-stat=14). The last feature is very broad and might be caused by background contamination, but the first two features are located at low wavelengths where source counts clearly dominate the background. There are potential absorption features at 18.5 and 19.5 \AA\ with strengths of about $\Delta$C-stat=11.

We perform 2009 Monte Carlo simulations to quantify the significance of these spectral residuals. We focus on the 3 strongest emission lines, each of them with at least $\Delta$C-stat=13. We find 97 simulated emission features stronger than this threshold, so the confidence of the weakest of our features is $\sim$95 \% each. Furthermore, we find 13 emission features with $\Delta$C-stat>16.7, which is the strength of the most prominent emission residual in real spectrum (confidence level of 99.35 \%). Finally, we want to quantify the confidence on multiple spectral features present in a fake spectrum with $\Delta$C-stat>13 (as a lower limit of confidence on our emission residuals). We find 11 occurrences of two lines in a single fake spectrum with such strength, but only 1 case of two emission lines which gives a confidence level of roughly 99.9 \% (even though we are definitely still in the discrete regime here given the total number of simulations). Therefore it seems extremely unlikely that all 3 emission features observed in the spectrum of NGC 5204 X-1 are caused by noise.

\begin{figure}
	\includegraphics[width=9cm]{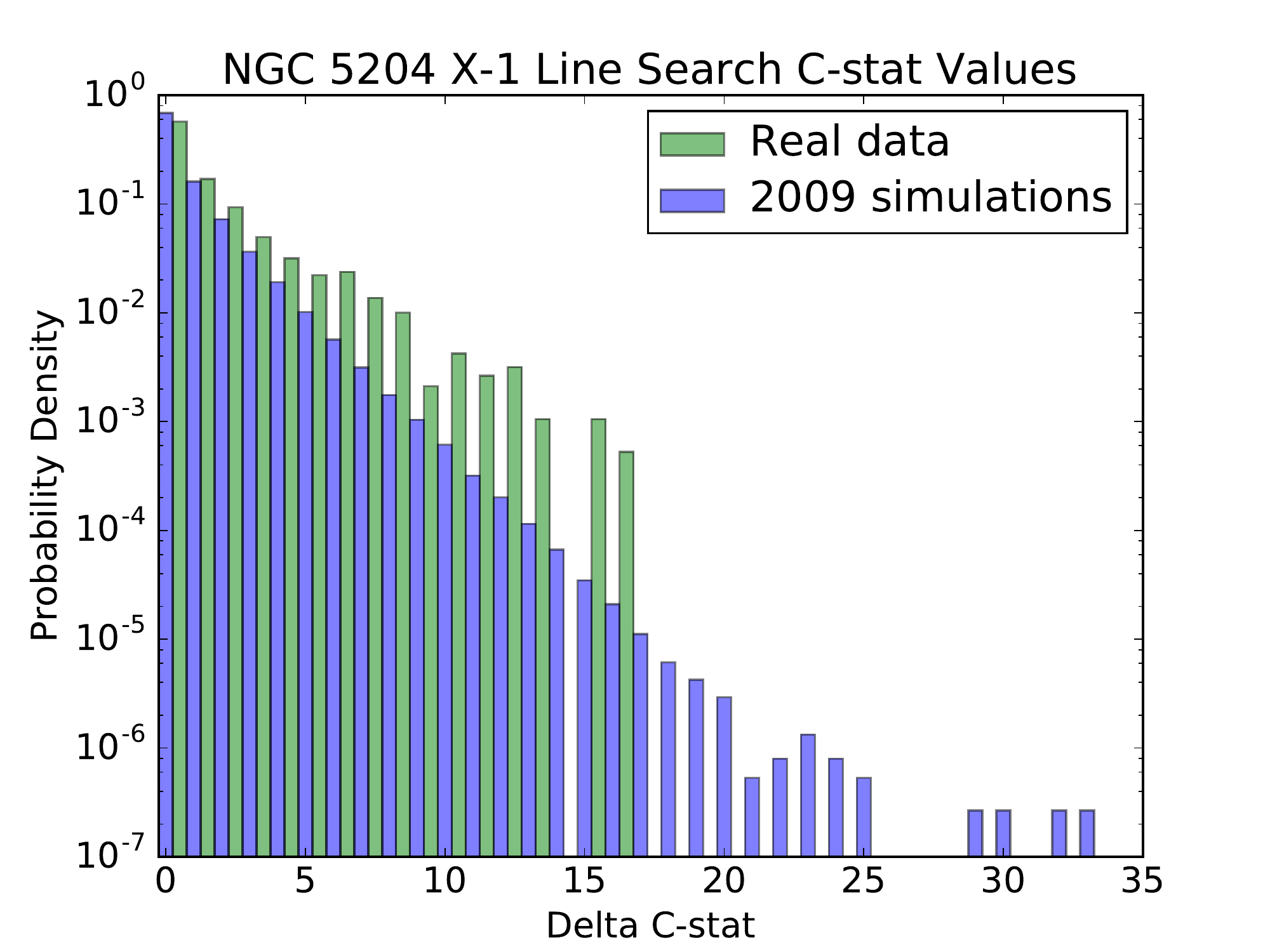}
    \caption{Histogram of $\Delta$C-stat of a real Gaussian line scan of NGC 5204 X-1 (green) and of line scans of 2009 Monte Carlo simulated datasets (blue). The Y axis is the probability density function, the integral of which is 1, and the X axis is the $\Delta$C-stat value.}
    \label{N5204_histogram}
\end{figure}

The significance of our detections can also be shown in a different way. We can plot a histogram of $\Delta$C-stat values (the fit improvement) using all energy bins in our band for a certain Gaussian line scan. A bin of $\Delta$C-stat=X is then equal to the number of occurrences of $\Delta$C-stat between X and X+1 in this line scan. We can also rescale the bin values to obtain the probability density of $\Delta$C-stat - in our case, such thing is achieved by simply dividing all bin values by 2000 (the number of energy bins). This is plotted in Fig. \ref{N5204_histogram}. In green, the histogram of the real Gaussian line scan of NGC 5204 X-1 (with a dispersion velocity of 1000 km/s) versus the $\Delta$C-stat value is shown, while the average histogram of line scans on 2009 Monte Carlo simulated spectra of NGC 5204 X-1 is in blue. The y axis in the histogram is the probability density of the $\Delta$C-stat value (the integral of which is 1). Note that the real data histogram does not contain a value at $\Delta$C-stat=14 - this simply means there were no occurrences of $\Delta$C-stat between 14 and 15 in the line scan. One can notice that the probability density of simulated data resembles very much a powerlaw function. In comparison, there is clear excess of higher $\Delta$C-stat data in our real search. One should however keep in mind that the last bin of real data (in green) corresponds to exactly 1 occurrence (hence 1 case $\sim0.5*10^{-3}$ in the histogram) so the last few bins are affected by small number statistics. It seems very unlikely that the real data line scan distribution of $\Delta$C-stat values comes purely from noise which creates the blue histogram.

\subsection{NGC 6946 X-1}

NGC 6946 X-1 was already studied in \citet{Pinto+16}, but without a rigorous Gaussian line scan, hence we include it in our analysis. It is the softest ULX in the sample. At a distance of 5.6 Mpc, its PN count rate is only about 0.36 count/s, but luckily it has a full orbit 120 ks exposure (110 ks of clean data). We can fit it with a standard double blackbody plus a powerlaw model with temperatures T$_{BB} = 0.17 \pm 0.09$ keV and T$_{MBB} = 1.68^{+0.13}_{-0.20}$ keV and a gamma index of $2.7 \pm 0.4$, although the fit is not very good at C-stat=195 with 118 degrees of freedom mostly due to a prominent residual at around 1 keV \citep{Middleton+14}.

\begin{figure*}
	\includegraphics[width=18cm]{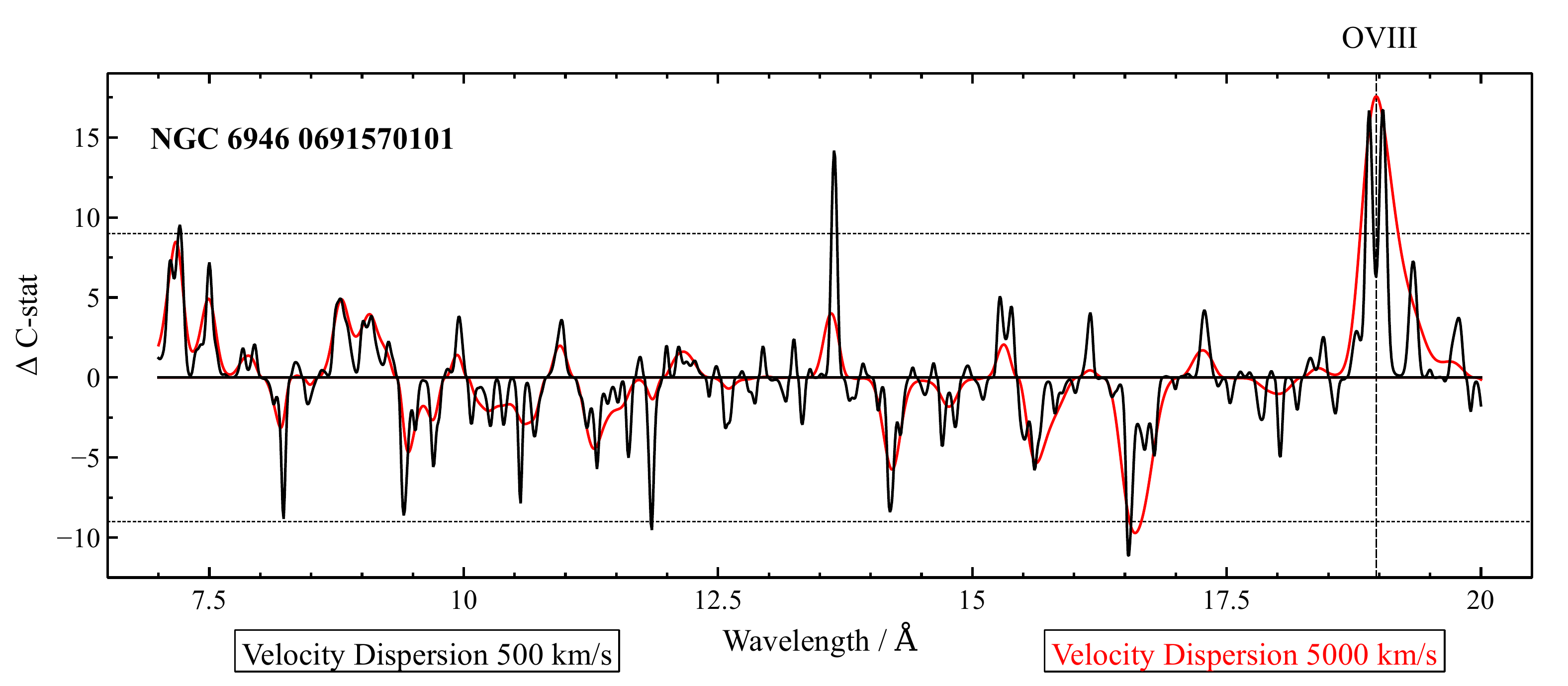}
    \caption{Line search results for NGC 6946 X-1. Axes are defined as in Fig. \ref{HoIX}.}
    \label{N6946}
\end{figure*}

The observation has about 3000 RGS source counts in total. We had to use a custom background since one of the 2 default background regions was contaminated by multiple bright X-ray binaries within the host galaxy. We use the PN spectral model which we renormalise as we did with other sources and search for spectral lines between 7 and 20 \AA. The line search finds a narrow emission feature at 13.5 \AA\ at $\Delta$C-stat=14 and a very strong and broad emission feature at 19 \AA\ with a strength of over 17.5 $\Delta$C-stat. There is also an absorption residual at 16.5 \AA\ with $\Delta$C-stat of about 10 and other potential weaker features. The Gaussian line scan results are shown in Fig. \ref{N6946}.

We perform 1070 Monte Carlo simulations and Gaussian line scan them. We focus on the prominent signature at 19.3 \AA\, with a width of about 3000 km/s. In total we find 4 fake spectral features stronger than the real feature. This suggests the feature is significant at about 3$\sigma$, but further simulations would be necessary to constrain its significance more accurately.

\subsection{NGC 4631 ULX1 and IC 342 X-1}

A thorough search on the XSA archive shows that also NGC 4631 ULX1 and IC 342 X-1 have suitable pointings for RGS analysis. Unfortunately, NGC 4631 ULX1 is relatively distant at over 7 Mpc and has a flux of only about 0.1 count/s with PN. Even though its observation has a relatively long exposure of 55 ks, this only results in about 650 RGS counts in total. We consider these data to be too low in quality to be able to detect any spectral features because the source flux is at the same level as the background flux.

IC 342 X-1 has 2 observations with good enough pointing for RGS with a total exposure of 120 ks and a sufficient flux of $\sim$0.5 count/s with PN. Unfortunately, the RGS spectrum is contaminated by another source with $\sim$half the X-ray luminosity (in the 0.3 to 10 keV band) of IC 342 X-1 along the dispersion direction of the spectrometer and hence it is unusable for subtle analysis like a spectral line search.

\section{Discussion}

We have collected all available XMM RGS X-ray ULX data with good enough quality for spectral line searches. Then we searched for residuals in the continua of their spectra by performing Gaussian line scans. The statistical significance of some among the most prominent features was quantified by Monte Carlo simulations of source spectra.

\subsection{Strongest features}

Table \ref{linedata} shows the strongest residuals found in the RGS spectra of ULXs in our sample. This means their $\Delta$C-stat difference is higher than 9 for at least one value of the line width (we have searched using the line width equivalents of 500 km/s, 1000 km/s and 5000 km/s). We found that there is usually little difference between the results from searches with the line widths of 500 and 1000 km/s.

\subsection{Attempting to identify the spectral features}

\begin{table*}
	\centering
	\caption{The most prominent residuals found by the Gaussian line search, ordered by the source hardness ratio as defined earlier. Columns (1) and (2) list the source name and the observation used for line scan. Columns (3) and (4) show the wavelength of the residual in \AA\ and whether it is an absorption or emission feature. Columns (5) and (6) contain the $\Delta$C-stat value using 2 different line widths, i.e. using 2 different gas dispersion velocities. Column (7) lists the preliminary identification of a residual, and column (8) shows the statistical significance of the feature obtained by Monte Carlo simulations.}
	\label{linedata}
	\small\addtolength{\tabcolsep}{-1.5pt}
	\begin{tabular}{cccccccc}
		\hline
		Object&Observation&Wavelength&Line&$\Delta$C-stat&$\Delta$C-stat&Preliminary&Significance\\
		 & & \AA\ & type & (500 km/s) & (5000 km/s) & identification & \% \\
		(1)  & (2) & (3) & (4) & (5) & (6) & (7) & (8) \\
		\hline
		NGC 5643 X-1 & 0744050101 & 13.79 & Em. & 11.43 & 6.42 &  &  \\		
		%Holmberg IX X-1 & 0200980101 & 8.40 & Em. & 10.29 & 8.61 & Fe XXIII (2s2-2s.4p) / Fe XXIV (2p-4d) / Mg XII (1s-2p) &  \\
		Holmberg IX X-1 & 0200980101 & 8.40 & Em. & 10.29 & 8.61 & Mg XII (1s-2p) &  \\
		Holmberg IX X-1 & 0200980101 & 12.48 & Abs. & -9.56 & -9.45 &  & $\sim$98 \\
		Holmberg IX X-1 & Simultaneous & 8.63 & Em. & 11.52 & 10.78 & Mg XII (1s-2p) &  \\
		NGC 4190 ULX1 & Simultaneous & 17.07 & Em. & 9.04 & 6.95 & Fe XVII (2p-3s) &  \\
		NGC 4190 ULX1 & Simultaneous & 18.62 & Em. & 9.01 & 12.20 & O VII (1s-3p) / O VIII (1s-2p) &  \\
		M33 X-8 & Simultaneous & 12.57 & Em. & 10.73 & 9.69 & &  \\
		NGC 1313 X-2 & Simultaneous & 18.94 & Em. & 10.50 & 7.15 & O VIII (1s-2p) &  \\
		NGC 1313 X-2 & Simultaneous & 8.99 & Abs. & -9.16 & -9.78 &  &  \\
		NGC 1313 X-2 & Simultaneous & 9.58 & Abs. & -13.90 & -10.58 &  &  \\
		NGC 1313 X-2 & Simultaneous & 12.98 & Abs. & -11.42 & -4.84 &  &  \\
		NGC 5204 X-1 & Stack & 9.97 & Em. & 16.70 & 13.07 &  & $\gtrsim$99.35 \\
		NGC 5204 X-1 & Stack & 11.33 & Em. & 12.85 & 7.87 &  & 95 \\
		NGC 5204 X-1 & Stack & 18.44 & Abs. & -11.18 & -2.4 &  &  \\
		NGC 5204 X-1 & Stack & 19.67 & Abs. & -9.88 & -11.19 &  &  \\
		NGC 5204 X-1 & Stack & 20.55 & Abs. & -9.24 & -3.48 &  &  \\
		NGC 5204 X-1 & Stack & 22.04 & Em. & 13.04 & 13.99 & O VII triplet & 95 \\
		Holmberg II X-1 & 0200470101 & 11.79 & Em. & 11.18 & 7.21 & Fe XXII (2p-3d) / Fe XXIII (2p-3d) & \\
		Holmberg II X-1 & 0200470101 & 14.67 & Abs. & -12.18 & -3.80 &  &  \\
		Holmberg II X-1 & 0200470101 & 21.97 & Em. & 12.20 & 3.61 & O VII triplet & 93  \\
		Holmberg II X-1 & 0200470101 & 22.26 & Abs. & -10.28 & -0.24 &  &  \\
		Holmberg II X-1 & Simultaneous & 11.79 & Em. & 11.36 & 6.88 & Fe XXII (2p-3d) / Fe XXIII (2p-3d) & \\
		Holmberg II X-1 & Simultaneous & 21.96 & Em. & 15.40 & 14.42 & O VII triplet &  \\
		NGC 6946 X-1 & 0691570101 & 7.22 & Em. & 9.54 & 8.50 &  &  \\
		NGC 6946 X-1 & 0691570101 & 11.86 & Abs. & -9.53 & -1.36 &  &  \\
		NGC 6946 X-1 & 0691570101 & 13.65 & Em. & 14.18 & 4.01 &  &  \\
		NGC 6946 X-1 & 0691570101 & 16.55 & Abs. & -11.13 & 9.72 &  &  \\
		NGC 6946 X-1 & 0691570101 & 18.98 & Em. & 16.69 & 17.57 & O VIII (1s-2p) & $\gtrsim$99.7\\

		\hline
	\end{tabular}
\end{table*}

In NGC 6946 X-1, we detect an emission residual located at the rest-frame wavelength of oxygen VIII (19.0 \AA) with a significance of at least 3$\sigma$. The line is moderately broad with a width of 0.2 \AA\ ($\sim$3000 km/s).

In NGC 5204 X-1, we find 3 emission features. A broad emission line at $\sim$22 \AA\ (approximate rest-frame wavelength of the O VII triplet) and a significance of 95 \%. The remaining 2 emission lines do not correspond to a rest-frame wavelength of any expected (i.e. strong enough) elemental transition from a photoionized plasma or plasma in collisional equilibrium. However, they are most likely not resulting from noise: the significance of the stronger one (at 10 \AA) is about 99.35 \%, and of the weaker one (at 11.3 \AA) is $\sim$95 \%. These 2 features were already noticed by \citet{Roberts+06} in Chandra data, although due to poor spectral resolution, they blend into a single broad emission line in CCD spectra. The fact that the lines are not located at the rest-frame wavelength of a transition makes their identification much more challenging. We manually experimented with photoionization models such as \textsc{pion} and collisional equilibrium models like \textsc{cie} to represent emitters at different blueshifts or redshifts but were unable to identify the lines this way without an automated approach (see Sect. \ref{spectral_model_search}).

Holmberg II X-1 shows 2 strong emission residuals, both in the highest quality observation (0200470101) and in the simultaneous analysis of 2 observations at once. The residuals are located at the rest-frame wavelengths of iron and oxygen: the first one corresponds to Fe XXII and/or Fe XXIII at $\sim$11.8 \AA, and the second one is at the rest-frame wavelength of the O VII triplet. Both features are narrow in the single observation, while the O VII feature becomes wide if we include all observations available. The long observation also shows multiple absorption features, with a combined significance of the strongest 2 being about 3$\sigma$ based on Monte Carlo simulations. These features are much weaker in the simultaneous line scan.

Holmberg IX X-1 seems to possess only 1 strong emission feature. In the highest quality observation (0200980101) search, this feature is centred on the rest-frame wavelength of Mg XII (8.42 \AA). Curiously, in the simultaneous search with all observations available, the feature is much weaker but there is a strong emission line shifted by 0.23 \AA\ ($\sim$8000 km/s) at 8.63 \AA. This could either be the same feature observed in the long observation or a completely different signature. The one notable absorption feature is only present in the long observation search at 12.48 \AA. It is a broad feature, resembling the absorption lines found in NGC 1313 X-1, albeit at a different velocity shift. If this is a signature of an outflow, it could suggest a different outflow velocity or simply a different viewing angle of the ULX and its wind. Monte Carlo simulations show we are detecting this feature at about 98 \% confidence level.

\subsection{Physical model search}

Where the spectral residuals are not located at a rest-frame wavelength of any expected elemental transition, we need to use SPEX models to identify any redshifted or blueshifted lines (for more details, see Sect. \ref{spectral_model_search}).

We follow-up with a search for spectral features in Holmberg IX X-1 using an ionised absorber model. We adopt a grid of velocities from 0 to -100000 km/s with a step of 500 km/s (which is comparable to the RGS resolution). Then we generated a \textsc{xabs} model in SPEX (a photoionized absorber model). We adopt a velocity broadening in \textsc{xabs} of 150 km/s (parameter "v" in \textsc{xabs} model, then do the same for 750 and 1500 km/s). We update the velocity shift of the \textsc{xabs} model according to the velocity grid above (parameter "zv" in \textsc{xabs} model) in each step of the search. The spectrum is then fitted in SPEX, while leaving the column density N$_{H}$ and the ionisation $\xi$ parameter free. The $\Delta$C-stat value is calculated for every velocity shift, as well as errors on N$_{H}$ and $\xi$. We find the best fit is obtained for a blueshift of either 0.06c or 0.26c. The best fit improvement is about $\Delta$C-stat of 15, which is not a high enough significance to conclusively claim a wind detection. Further deep and uninterrupted exposures are needed for a firmer conclusion.

We apply the same procedure to the highest quality observation of Holmberg II X-1. Adopting a velocity dispersion of 500 km/s (based on the narrow shape of residuals found by line search) gives the best fit at a blueshift of $\sim$0.2-0.25c, albeit at a lower significance than obtained for Holmberg IX X-1. The best fit photoionization parameter in this case is log $\xi = 3.0 \pm 0.2$, similar to outflows that have been found in NGC 1313 X-1 and NGC 5408 X-1.

\subsubsection{A jet detection in NGC 5204 X-1?}

\begin{figure*}
	\includegraphics[width=18cm]{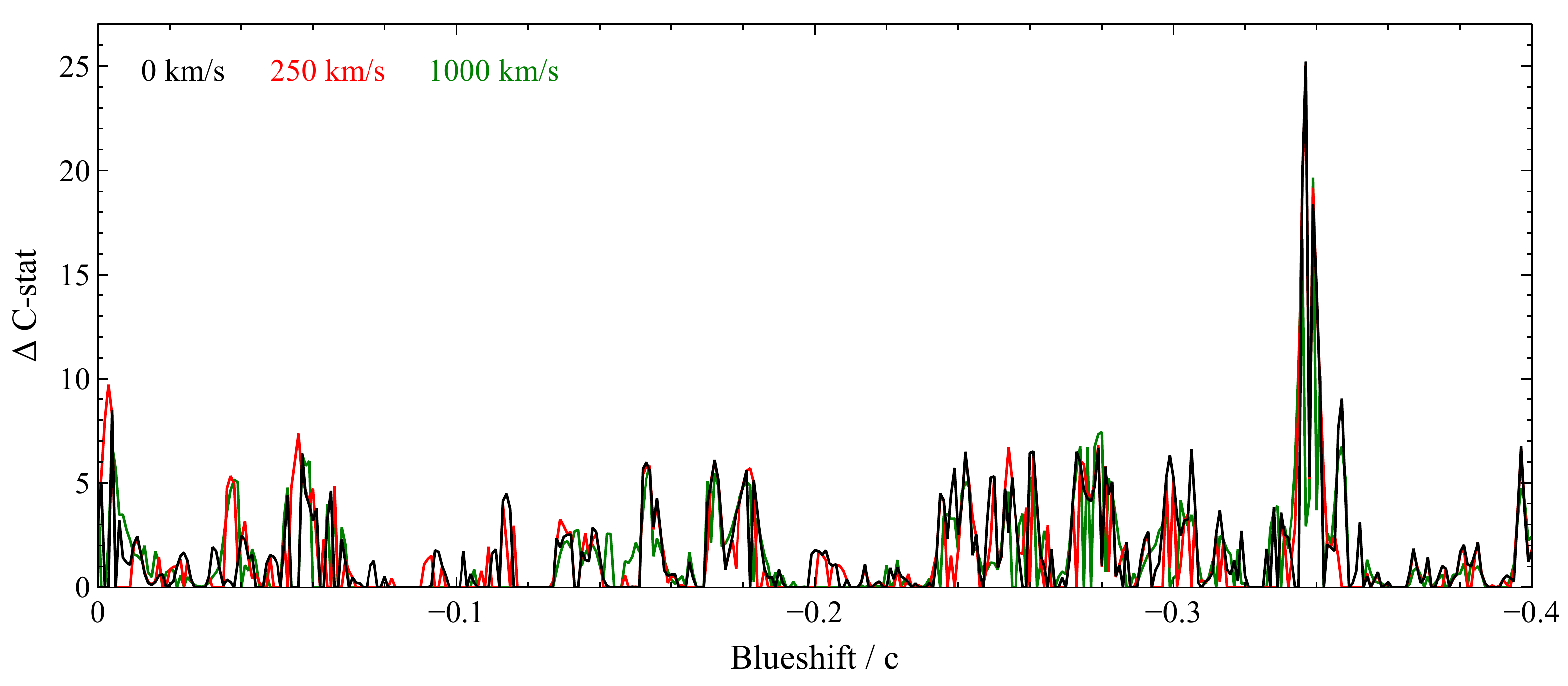}
    \caption{The \textsc{cie} scan results of the stacked NGC 5204 X-1 RGS spectrum for 3 different velocity broadening values.}
    \label{N5204_cie_scan}
\end{figure*}

\begin{figure*}
	\includegraphics[width=18cm]{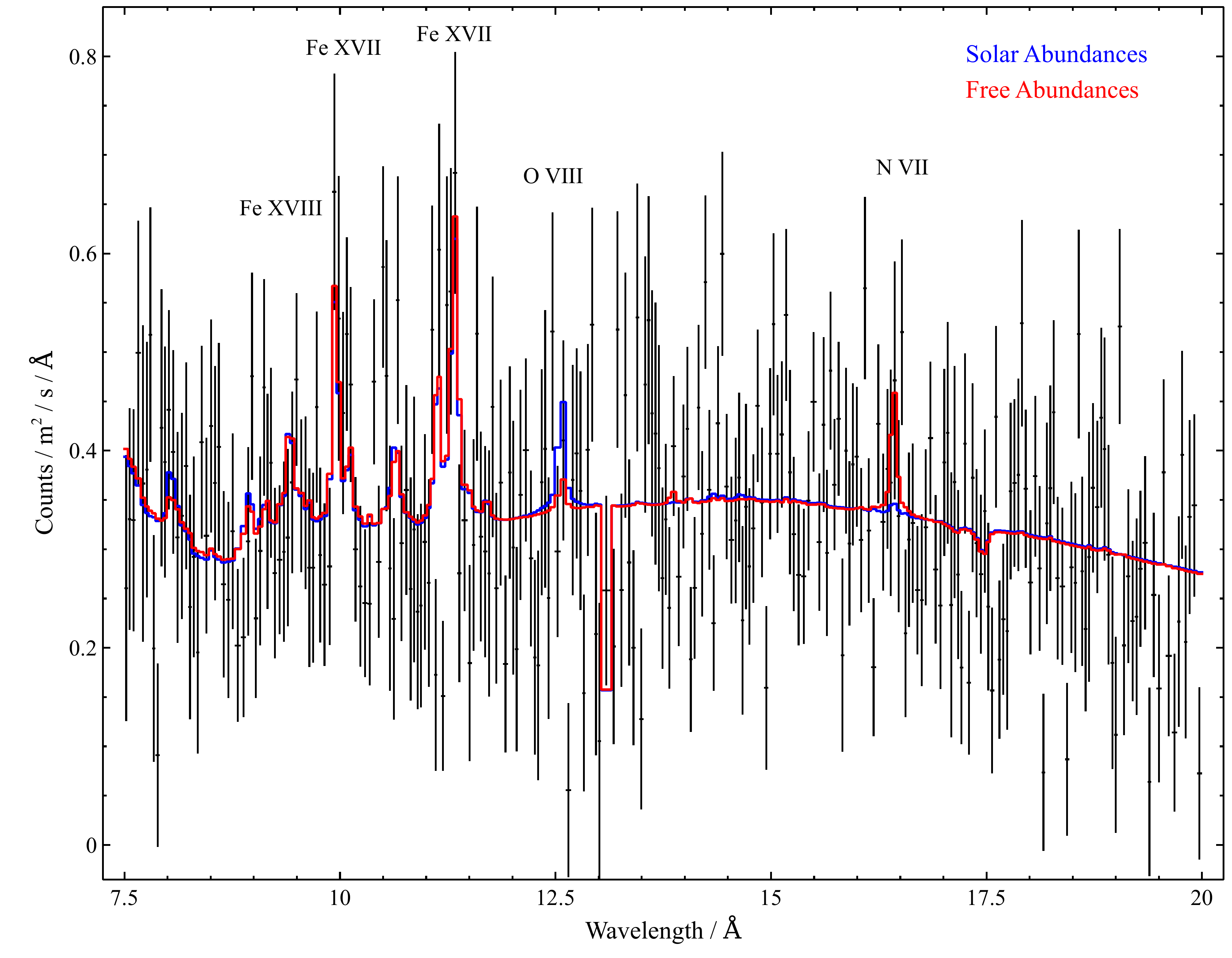}
    \caption{The RGS spectrum of NGC 5204 X-1 between 7.5 and 20 \AA, using all available XMM-Newton observations (stacked). The model is composed of a continuum fit with a powerlaw and a blackbody component (plus neutral absorption). On top of the continuum, a blueshifted collisionally ionised gas in equilibrium model \textsc{cie} is added. In blue, the model is shown with default abundances. In red, the abundances of nitrogen and oxygen are freed, and the abundances of neon and magnesium are tied to oxygen.}
    \label{N5204_cie_spectrum}
\end{figure*}

NGC 5204 X-1 shows multiple emission features, the most prominent ones being at 10 and 11.3 \AA. To explain them with a physical model, we perform a \textsc{cie} model scan between 0 and 0.4c blueshift on the stacked data. The results of the scan are shown in Fig. \ref{N5204_cie_scan}. We find a significant fit improvement for an outflow velocity of -0.337c with $\Delta$C-stat=22.08, a temperature of about 0.5 keV and a dispersion velocity of 1000 km/s. To follow up, we fit the RGS spectrum directly with a blueshifted \textsc{cie} model plus continuum. The fit is shown in Fig. \ref{N5204_cie_spectrum}, now on a narrower band between 7.5 and 20 \AA, which is less affected by the background. The collisionally ionised plasma model fits both emission features very well. These features are produced by Fe XVII in our model. The fit improvement is $\Delta$C-stat=25.87 for 4 degrees of freedom (normalisation, temperature and velocity dispersion of the \textsc{cie} component, plus its blueshift). The best fit is achieved for a velocity broadening of $\lesssim$1000 km/s.

The fit can be further improved by freeing some of the abundances. Since the \textsc{cie} component is mostly driven by the two iron XVII lines (and not continuum collisionally ionised emission from hydrogen bremsstrahlung), it is impossible to obtain any reliable results if we free the iron abundance. We thaw nitrogen and oxygen abundances. Then we couple neon and magnesium abundances to that of oxygen as they are thought to have a similar core-collapse supernova origin \citep{deplaa+07}, and it is not recommended to fit all elements separately due to low statistics. In the end, the addition of the \textsc{cie} component improves the fit by $\Delta$C-stat of 39.2 for 6 additional degrees of freedom - normalisation, temperature, velocity broadening, blueshift, and 2 abundances. The temperature of emitting gas is $0.60^{+0.08}_{-0.09}$ keV and the 1$\sigma$ upper limit on its velocity broadening is 335 km/s. The nitrogen abundance (with respect to iron) N/Fe is rather high but poorly constrained at N/Fe $\gtrsim$ 4 and the coupled oxygen, neon and magnesium abundance O/Fe = Ne/Fe = Mg/Fe $\lesssim$ 0.8. The blueshift of the \textsc{cie} component is $-0.3371^{+0.0006}_{-0.0005}~c$ (116800 km/s).

We perform Monte Carlo simulations to quantify the significance of these features. We simulate a fake RGS spectrum with comparable statistics and continuum model as the NGC 5204 X-1 spectrum. Then we launch the same \textsc{cie} scan procedure as we used for real data. We repeat the same process as many times as possible and look for any cases where a fake feature produces a stronger ($\Delta$C-stat) fit improvement than we found in the real spectrum. We performed 2112 simulations in total and found 6 outliers stronger than the feature found in measured data. This gives the significance of 99.7 \%, which is about 3$\sigma$.

Our findings are very similar to X-ray lines of SS 433, the Galactic microquasar \citep{Margon+84}, albeit at much higher blueshift. Most of the SS 433 X-ray emission comes from plasma ionised by its jets, which produces prominent emission lines throughout the X-ray band \citep{Marshall+02}. The gas producing these lines is accelerated to 0.27c, which, accounting for projection effects, results into emission from the blue jet being blueshifted by about 0.08c, and the red jet being redshifted by $\sim$0.16c. Unfortunately, due to the higher blueshift of lines seen in NGC 5204 X-1, the elemental transitions seen in SS 433 (with Chandra gratings) are mostly blueshifted out of the RGS energy band (7 \AA\ and higher). The only line seen both in our spectrum and the \citep{Marshall+02} analysis of SS 433 is the Ne X Lya and Fe XXIII line with the rest frame wavelength of 12.134 \AA. In SS 433, the line is seen at 11.194 \AA\ originating from the blue jet, and should be at 8.05 \AA\ in NGC 5204 X-1. Emission residuals can indeed be seen around 8 \AA\ in its spectrum (see Fig. \ref{N5204_cie_spectrum}), but they are too weak to make any claims. Further residuals in NGC 5204 X-1 are seen at higher energies in CCD PN and Chandra data \citep{Roberts+06}.

The blueshift of the emission lines seen in NGC 5204 X-1 is higher than in SS 433 (0.34c plus any projection effects versus 0.27c), but the terminal jet velocity might be source dependent. The spatial orientation of both sources is also most likely quite different with SS 433 being seen practically edge-on, while NGC 5204 X-1 is probably at a much lower inclination angle (but higher inclination than, say, NGC 1313 X-1 or Holmberg IX X-1 due to its softer spectrum). The X-ray continuum of SS 433 is much harder, and not super-Eddington (SS 433 does not look like a ULX from our point of view) due to inclination and heavy absorption. Hence most of the continuum emission of SS 433 actually comes from its jets and not the accretion disk as we observe in ULXs. Therefore a hardness comparison of SS 433 with other ULXs does not make sense. The temperature of the plasma observed in NGC 5204 X-1 is rather low at $\sim$0.6 keV, which is at the lower limit of estimates from the line strengths in SS 433 (0.5-10 keV). However, the NGC 5204 X-1 plasma temperature could be an underestimate as we can only see a few lines (most importantly Fe XVII) that are not blueshifted out of the RGS energy band. Future Chandra grating or calorimeter observations will be able to determine whether the source has other prominent emission lines at higher energies, as observed in SS 433. It is worth mentioning that NGC 5204 X-1 and SS 433 also exhibit very similar He I (6678 \AA), He II (4686 \AA) and H$\alpha$ (6563 \AA) line emission in the optical band \citep{Fabrika+15}.

\subsection{ULX comparison within the sample and with previous work}

Holmberg II X-1 and NGC 5204 X-1 have similar spectral hardnesses (Table \ref{ulxdata}) so are good candidates for a comparison within our sample. Looking at their line scans, Fig. \ref{HoII} and \ref{N5204}, we notice that both sources have strong detections of O VII line at rest frame. We also do not see any O VIII detection. At lower wavelengths, the line scans show emission residuals, but they are not exactly at the same energy. In Holmberg II X-1, we find a strong emission feature at 11.8 \AA, which could be associated with an iron transition. NGC 5204 X-1 does not show any residuals at this wavelength. Instead, it has 2 prominent emission features at lower wavelengths (10 and 11.3 \AA), which are difficult to associate with any rest-frame transitions but are likely produced by blueshifted collisionally ionised gas. There are no obvious absorption features that these 2 sources would have in common.

Next we would like to compare our findings with the ULXs that have already been shown to possess outflows \citep{Pinto+16,Pinto+17}. There are 3 such sources in total, as of June 2017: NGC 1313 X-1, NGC 5408 X-1 and NGC 55 ULX.

\begin{figure*}
	\includegraphics[width=18cm]{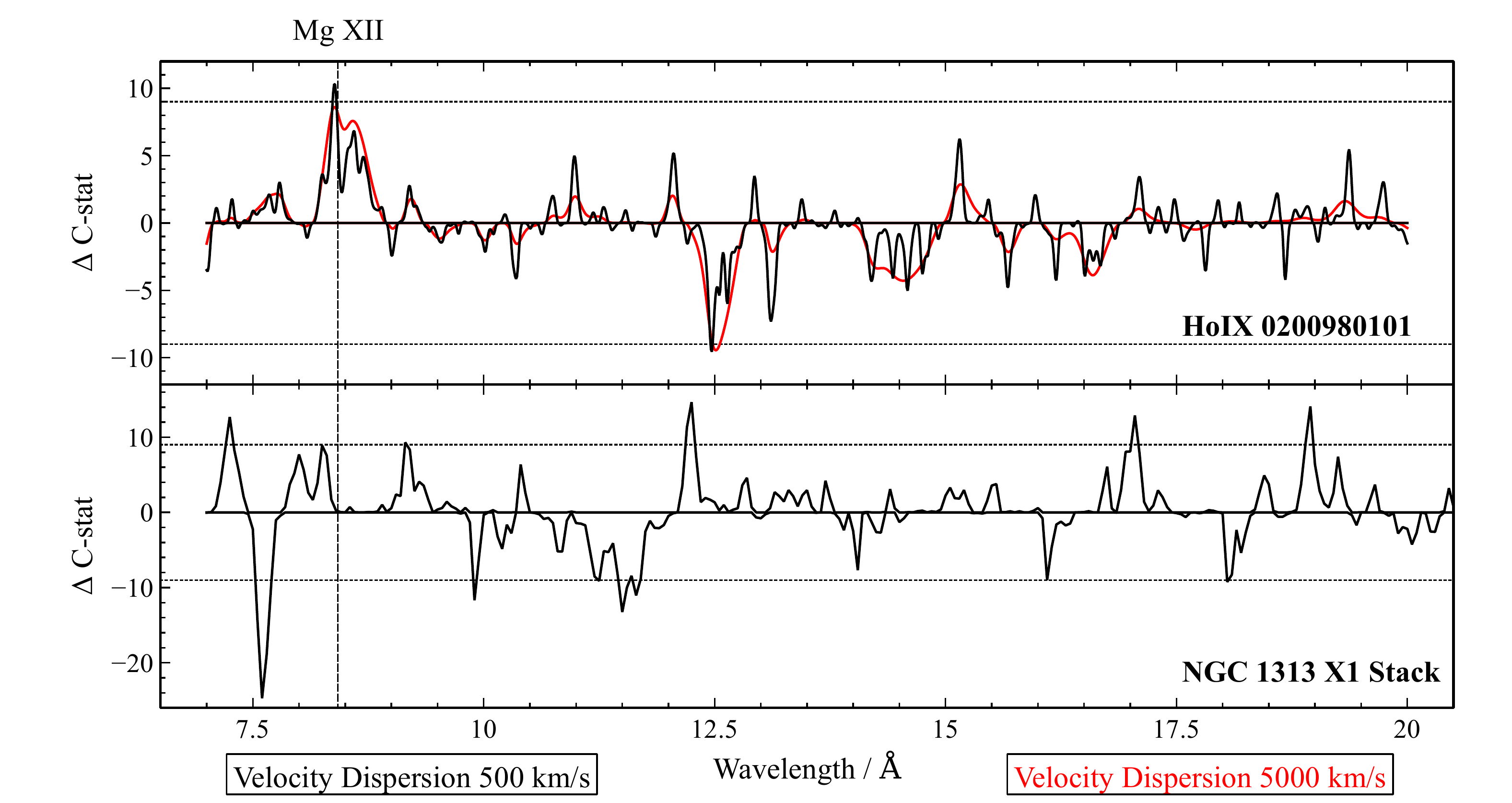}
    \caption{Comparison between a Gaussian line search of the highest quality single observation of Holmberg IX X-1 and the stacked results of NGC 1313 X-1. Axes are defined as in Fig. \ref{HoIX}.}
    \label{HoIXvsN1313}
\end{figure*}

Holmberg IX X-1 (average hardness ratio of 0.6) is one of the hardest sources in our sample (Table \ref{ulxdata}). Its spectrum resembles the one of NGC 1313 X-1 (hardness ratio 0.49), which was the first source with an outflow that was identified thanks to its shifted, broad absorption signature at around 11.5 \AA\ (Fig. \ref{HoIXvsN1313}). Holmberg IX X-1 does not have such a prominent absorption line, however it does show a broad residual at 12.5 \AA. If these 2 features are a result of the same transition, the projected velocity difference of outflows in these 2 sources needs to be about $\sim$0.08c (the one in NGC 1313 X-1 has a velocity of about 0.2c). Other than this, the line scans of these 2 sources do not have much in common.

\begin{figure*}
	\includegraphics[width=18cm]{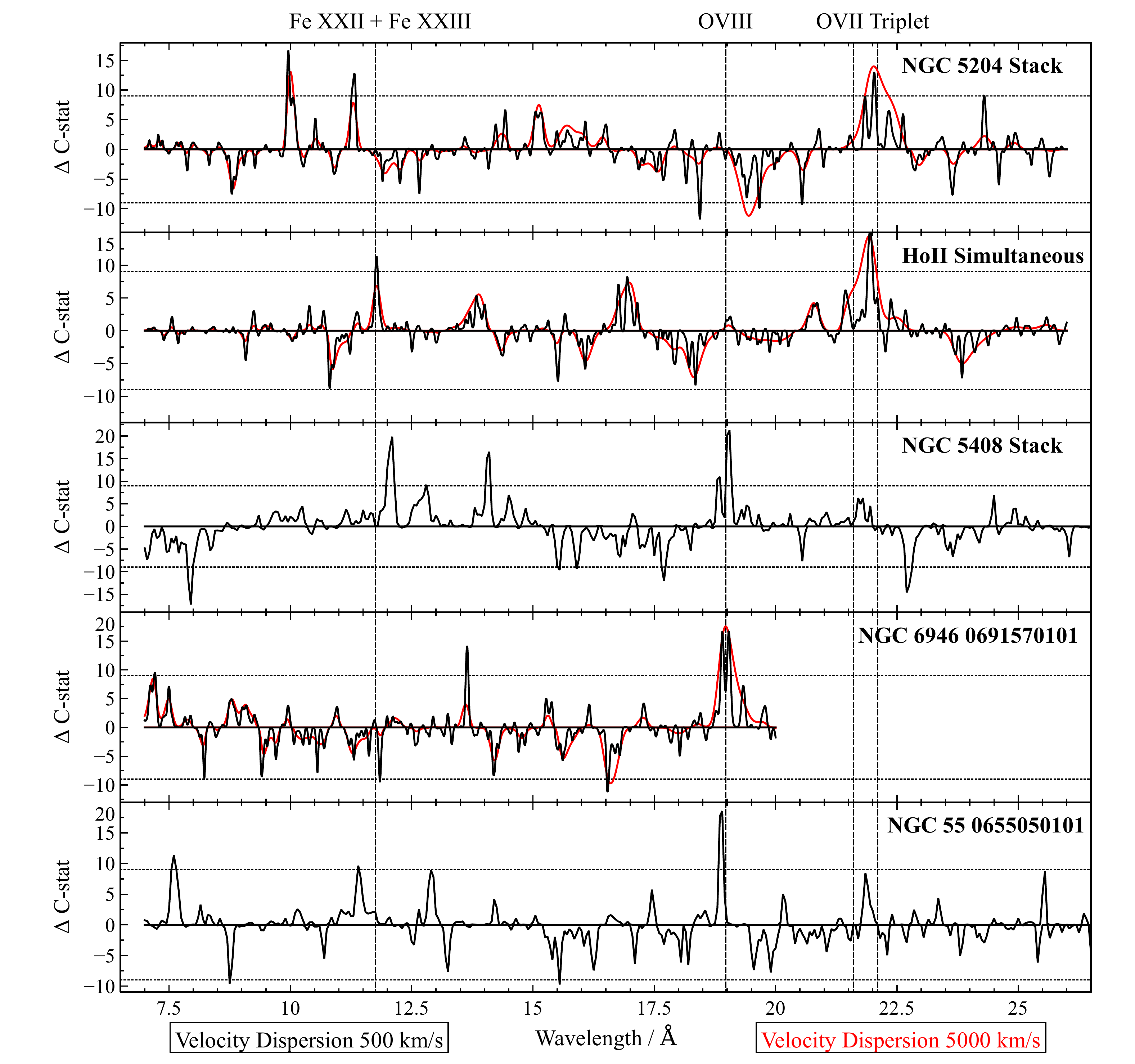}
    \caption{Comparison between a Gaussian line search of the stacked scan of NGC 5204 X-1, the simultaneous scan of Holmberg II X-1, the stacked results of NGC 5408 X-1, and results from single observations of NGC 6946 X-1 and NGC 55 ULX. Axes are defined as in Fig. \ref{HoIX}.}
    \label{HoIIvsN5408vsN5204vsN6946vsN55}
\end{figure*}

NGC 5408 X-1 (hardness ratio 0.19) has a softer X-ray spectrum than NGC 1313 X-1 or Holmberg IX X-1, similar to Holmberg II X-1 and NGC 6946 X-1 (both with HR $\sim$ 0.24) and (less similar) to the spectrum of NGC 5204 X-1 (HR = 0.4). Comparing these 4 sources (Fig. \ref{HoIIvsN5408vsN5204vsN6946vsN55}), we find their line scans give quite different results. While both NGC 5204 X-1 and Holmberg II X-1 show strong emission residuals at the O VII transition (22 \AA), there seems to be only a very small hint of O VII in the line scan of NGC 5408 X-1. Conversely, where NGC 5408 X-1 and NGC 6946 X-1 show a prominent emission line of O VIII (19 \AA), NGC 5204 X-1 and Holmberg II X-1 have little or no residuals. At lower wavelengths, most of these sources exhibit emission residuals in the 10 to 15 \AA\ range, but they are all shifted differently in each ULX. Holmberg II X-1 has one of the emission features exactly at the rest-frame wavelength of Fe XXII and Fe XXIII, and the 12.1 \AA\ feature in NGC 5408 X-1 can be explained by the Ne X transition. The other residuals do not seem to align with any expected rest-frame elemental transitions. However, the emission features of these 3 ULXs look curiously similar - they are at $9.954^{+0.018}_{-0.003}$ and $11.33^{+0.04}_{-0.02}$ \AA\ in NGC 5204 X-1, at $11.78\pm0.02$ and $13.87^{+0.08}_{-0.21}$ \AA\ in Holmberg II X-1 and at $12.07\pm0.03$ and $14.09^{+0.02}_{-0.03}$ \AA\ in NGC 5408 X-1 (errors obtained by fitting Gaussian lines to ULX spectra). Unfortunately, despite the similarity, all sources have slightly different relative ratios of wavelengths of their residuals: $0.881^{+0.003}_{-0.004}$ for NGC 5204 X-1, $0.849^{+0.015}_{-0.007}$ for Holmberg II X-1 and $0.857\pm0.004$ for NGC 5408 X-1. The line ratios of Holmberg II X-1 and NGC 5408 X-1 are consistent, but mostly because the 13.9 \AA\ feature in Holmberg II X-1 is rather weak and broad, hence the errors on its wavelength are large. The ratios of the first two objects are inconsistent with the line ratio of NGC 5204 X-1. Hence it seems very unlikely that the residuals correspond to identical elemental transitions in all 3 sources, (blue/red)shifted differently in each ULX as one could hope.

NGC 6946 X-1 has the softest spectrum from our ULX sample (Table \ref{ulxdata}), resembling that of the ULX/ULS source in NGC 55 (hardness ratio $\sim$ 0.11). In Fig. \ref{HoIIvsN5408vsN5204vsN6946vsN55}, we can see that both objects show a prominent O VIII emission line, which is blueshifted in NGC 55 ULX by a few hundred km/s. In the case of NGC 6946 X-1, the line is much wider but seems to be centred exactly on the rest-frame wavelength of O VIII. Additionally, both objects show other emission and absorption residuals, but none of them appear at the same wavelengths.

\subsection{Contamination by host galaxy}

Most of our sources have emission residuals right at the rest-frame wavelengths of oxygen VII and VIII transitions. The host galaxies of these ULXs emit in O VII and VIII and can potentially contaminate our results if the background subtraction is imperfect. This could be the case especially for galaxies with high star formation rates like NGC 6946, which are bright in this band ($<$1 keV). To check for such possibility, we take NGC 6946 X-1 as an example. We extract the source (background not subtracted) and the background spectrum, which is shown in Fig. \ref{N6946_bkg}. There is a small excess of flux around 19 \AA\ in the background spectrum, which is where the oxygen VIII transition is located, and also where we detect a significant emission residual in the ULX spectrum. However, the excess is too small to cause such a strong emission feature in our line scans even if partially underestimated (comparing with its strength in the NGC 6946 X-1 source spectrum). The source flux is about 2 times higher than the background flux in this energy band, hence it is unlikely that an imperfect subtraction of galaxy emission could produce such prominent spike in source spectrum.

We perform a further check to be sure that the oxygen features do not originate in galaxy emission only. We extract a MOS 1 detector image of the NGC 6946 X-1 pointing and use the rgsvprof\footnote{http://var.sron.nl/SPEX-doc/manual/manualse105.html} procedure to create a total flux profile along the RGS slit, in the RGS energy band from 0.35 to 1.77 keV. Based on this, we obtain the expected spectral broadening of a line created by such flux profile. The broadening is caused by sources off-axis in the dispersive spectrograph (in the wavelength direction) whose spectrum is then shifted in energy/wavelength. We create the flux profile for both source and background slits and compare them. Based on the line profiles, we estimate a 10 \%\ upper limit on galaxy contamination in the O VIII band.

\subsection{Further studies}

Our results show that with the current archival XMM-Newton data, we are able to tentatively detect narrow spectral features in most out of the 10 or so brightest ULXs in the sky. It also seems that, with one exception (NGC 5204 X-1), data available at the moment are insufficient in quality to distinguish between different emission and absorption models describing these features.

ULXs with well described outflows have been monitored for considerable amounts of time: over 400 ks of raw data in the case of NGC 1313 X-1 and $\sim$800 ks in the case  of NGC 5408 X-1. The current generation of X-ray observatories is able to make a difference for the brightest ULXs given enough observing time. Unfortunately, no other ULX have been observed for such long periods of time up to date, and most have total exposures of only about 100 ks (but usually not a full uninterrupted XMM-Newton orbit) or below.

It is clear that more data are necessary to put further constraints on the presence of outflows in the spectra of ULXs. At the moment, we struggle with 2 issues using the RGS XMM-Newton data: not enough counts in spectra for narrow line studies, and high background levels below 10 \AA\ and above 20 \AA. The first problem is solved purely by further exposures, the second one requires long, uninterrupted observations. The latter can be partially compensated by stacking separate observations, but potential long-term variability issues may compromise the results (especially for the most variable ULXs). It would be also useful to re-observe some objects (like IC 342 X-1), which have already been observed by XMM-Newton, but with a roll angle that makes the RGS analysis impossible, for example due to contamination by other sources.

\begin{figure}
	\includegraphics[width=8.4cm]{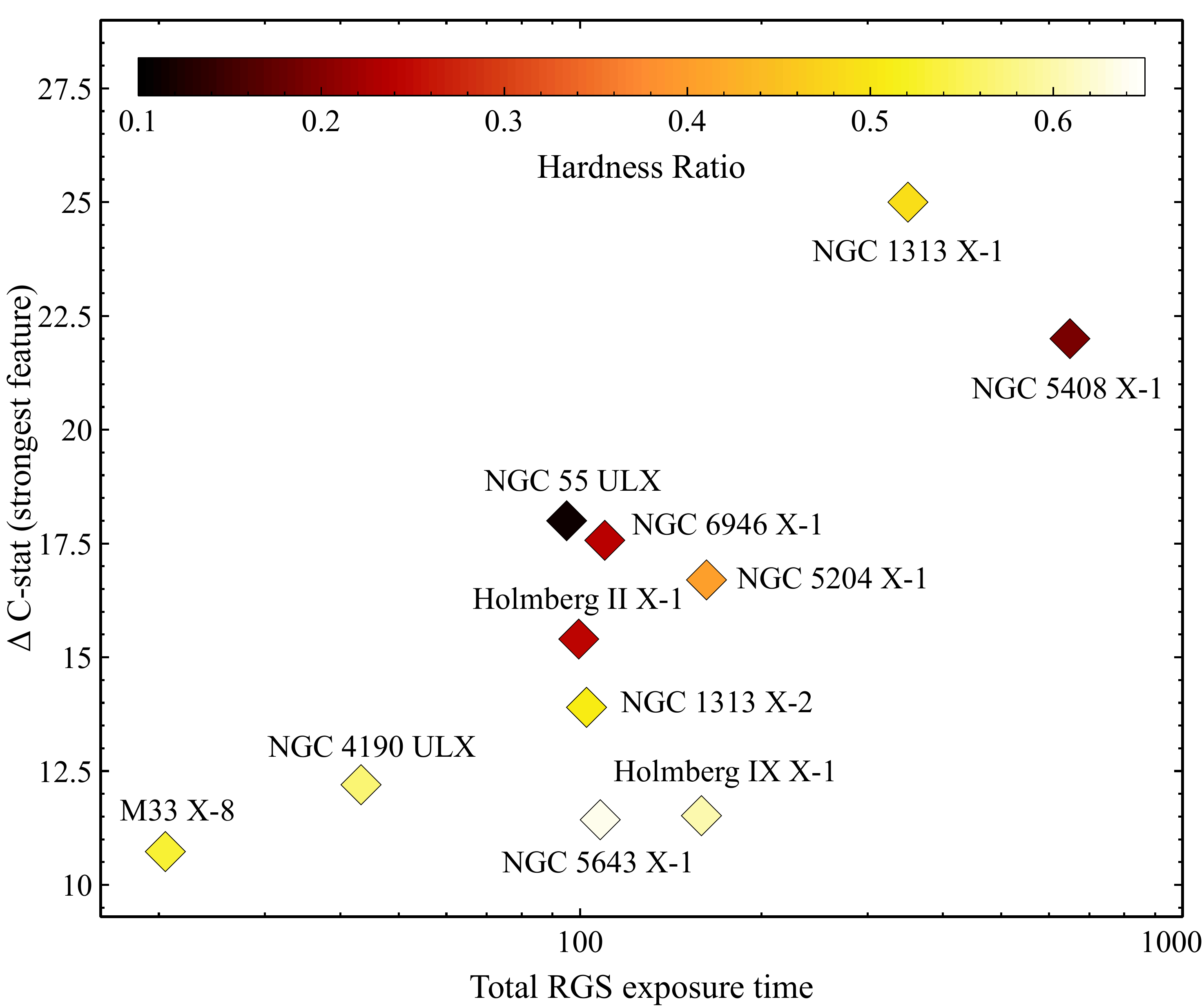}
    \caption{Plot showing the significance (in $\Delta$C-stat) of the strongest feature of a particular ULX versus the total clean RGS exposure time. Heat map shows the hardness ratio of the ULX (see Table \ref{ulxdata}). NGC 1313 X-1, NGC 5408 X-1 and NGC 55 ULX are not a part of our sample as they were studied by \citet{Pinto+16} and \citet{Pinto+17}.}
    \label{Significance_vs_time}
\end{figure}

Observational time aside, if we believe the funnel theory of the accretion in ULXs \citep[Fig. 13 of][]{Pinto+17}, some sources seem to be better candidates for a detection of outflow signatures than others. In Fig. \ref{Significance_vs_time}, the $\Delta$C-stat significance of the strongest spectral feature found by the Gaussian line scan is shown versus the total clean exposure time of the source. The color scheme here defines the hardness ratio of the source as defined in Table \ref{ulxdata}. There are many different factors that affect the detection significance of features other than just the exposure time, yet we can see that all the soft sources from our sample (NGC 6946 X-1, Holmberg II X-1 and NGC 5204 X-1) have strong detections of features, while the harder sources show weaker detections despite some of them having enough observing time, like Holmberg IX X-1 (or counts like M33 X-8). It is difficult to draw any firm conclusions from the plot (which needs to be more densely populated with new observations and other sources), but soft ULXs seem to be good candidates for future spectral line and outflow searches.

NGC 5204 X-1 is naturally the best source to start with. Our simulations show that a few hundred additional ks of XMM-Newton data should be sufficient to describe the observed spectral features to very high detail and significance. If the feature is indeed the blue side of a jet, the data should be of high enough quality to tentatively locate the red jet. This could be used to accurately determine the actual spatial orientation of the ULX. Chandra gratings, despite their lower collecting area, could be used to scan the higher energy band, where residuals (in CCD spectra) are seen as well, currently at lower statistical significance. Similar approach could be chosen for other promising sources. Most importantly, full-orbit XMM Newton observations (with proper alignment for the RGS detectors) will be crucial for further studies of ULX outflows, to both achieve the necessary total count amount, and constrain the background as well as possible.

A drawback of current instruments is that they both are dispersive spectrographs. Proper care is required when planning the observations to avoid contamination by other point sources in the field such as galactic binaries and AGNs. Additionally, in some cases the host galaxy contamination can also be an issue in the softest band ($<$0.7 keV), especially when it is difficult to obtain a representative background region that would accurately constrain the galactic emission.

Future missions such as XARM \citep[Hitomi replacement,][]{Takahashi+10} and Athena \citep{Nandra+13} will be able to overcome this difficulty thanks to their calorimeters, albeit in the harder X-ray band ($>$1 keV). Arcus\footnote{http://www.arcusxray.org/}, if approved, despite being a dispersive spectrograph, would prove crucial in characterising ULX outflows thanks to its unprecedented spectral resolution and collecting area in the soft X-ray band ($<$1.2 keV). These missions will naturally also require much less exposure time to achieve the expected results. They will also be capable of detecting outflows in ULXs at larger distances, thus increasing our sample size considerably.

\section{Conclusions}

We collected all the usable archival high spectral resolution data of ULXs. Aiming to find spectral features or directly outflows in ULX spectra, we performed Gaussian line scans, followed by Monte Carlo simulations of spectra and physical model scans for the most promising sources. We compared our results with the previous achievements in this field. Our results show that:  

\begin{itemize}

\item{In some sources, we discover several potential lines located at similar wavelengths to the rest-frame positions of the strongest lines of magnesium, iron and oxygen.}

\item{We find multiple strong emission residuals in the spectrum of NGC 5204 X-1. Most of these can be described by collisionally ionised plasma blueshifted to -0.34c. The significance of this detection is at least 3$\sigma$. The detected features resemble the X-ray line emission from the Galactic microquasar SS 433.}

\item{Holmberg IX X-1 exhibits 2 interesting features. An emission residual at around 8.5 \AA, which could correspond to rest-frame Mg XII emission, and an absorption residual at around 12.5 \AA. We were able to fit the spectrum with a photoionized absorption model with outflow velocity of $\sim$0.25c.}

\item{Holmberg II X-1 shows both absorption and emission residuals. The absorption features are only present in the highest quality observation, and when combined they are significant at about 3$\sigma$. The two emission features are present in both observations and might correspond to iron (Fe XXII and/or Fe XXIII) and O VII emission.}

\item{The spectrum of NGC 6946 X-1 shows a broad emission feature at 19 \AA, the rest-frame wavelength of O VIII transition. The line is detected at over 3$\sigma$ in high resolution data. }

\item{All remaining sources show moderately strong emission or absorption residuals. However, at this stage we prefer not to claim identification of these features with outflows as more Monte Carlo simulations and new data are required.}

\item{At the moment, we are limited by the data quality, more specifically by high background and low count number. Further uninterrupted and long observations are required to overcome both of these limitations. Full-orbit XMM observations and deep Chandra data will be crucial in further studies to discover and study ULX outflows in more detail. Future missions such as XARM, Arcus and Athena will also be able to achieve much better results with considerably less exposure time and for a larger sample of ULXs.}

\end{itemize}

\section*{Acknowledgements}

PK acknowledges support from the STFC. CP and ACF acknowledge support from ERC Advanced Grant Feedback 340442. DJW acknowledges support from STFC Ernest Rutherford fellowships (grant ST/J003697/2). This work is based on observations obtained with XMM-Newton, an ESA science mission funded by ESA Member States and USA (NASA). This research has made use of the NASA/IPAC Extragalactic Database (NED) which is operated by the Jet Propulsion Laboratory, California Institute of Technology, under contract with the National Aeronautics and Space Administration. This research has made use of the SIMBAD database, operated at CDS, Strasbourg, France.

%%%%%%%%%%%%%%%%%%%%%%%%%%%%%%%%%%%%%%%%%%%%%%%%%%



\bibliographystyle{mnras}
\bibliography{References} % if your bibtex file is called example.bib

%%%%%%%%%%%%%%%%%%%%%%%%%%%%%%%%%%%%%%%%%%%%%%%%%%

%%%%%%%%%%%%%%%%% APPENDICES %%%%%%%%%%%%%%%%%%%%%

\appendix

\section{Complete Gaussian Line Search Results}
\label{RGS_plot}

In this appendix we plot all the results from the RGS Gaussian line scan analysis. All ULXs with usable RGS data are shown.

\begin{figure*}
	\includegraphics[width=17.7cm]{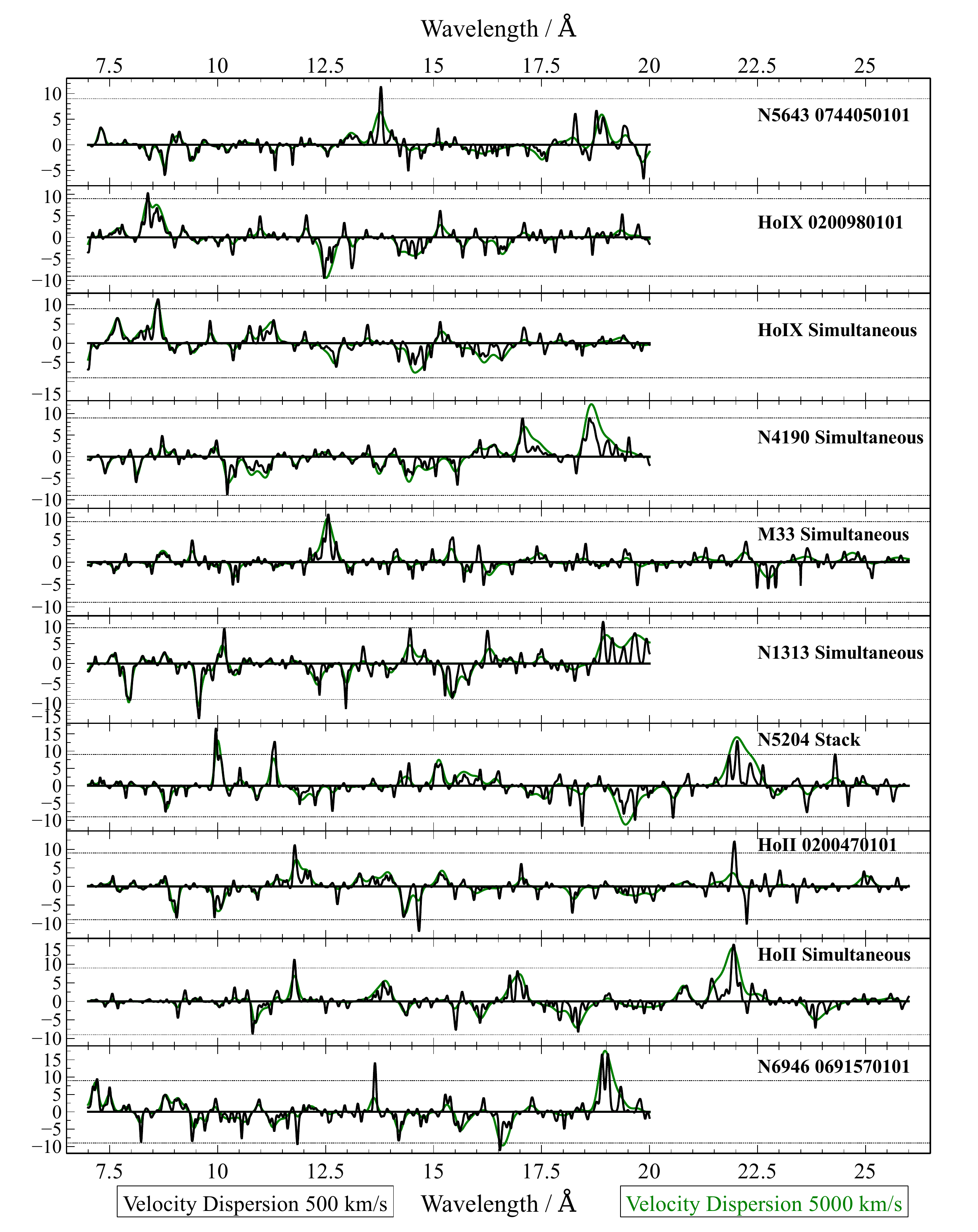}
    \caption{The Gaussian line search performed on the RGS spectra of ULXs in our sample. The wavelength is on the X-axis and the obtained $\Delta$C-stat value on the Y axis for each source or observation analysed. The objects are ordered by their hardness the same way as in Fig. \ref{fig:PN_all}, starting from the hardest source, NGC 5643 X-1, ending with the softest source, NGC 6946 X-1. The Y axis is defined as $\Delta$C-stat times the sign of the normalisation of the line to show the difference between absorption and emission residuals.}
    \label{RGS_all}
\end{figure*}

\section{NGC 6946 X-1 Source and Background Spectrum}

The appendix contains a plot of the source (not background-subtracted) spectrum of NGC 6946 X-1, using both RGS detectors, and the background spectrum used to study the source.

\begin{figure*}
	%\includegraphics[width=17.7cm]{N6946_bkg.pdf}
	\includegraphics[width=17.7cm]{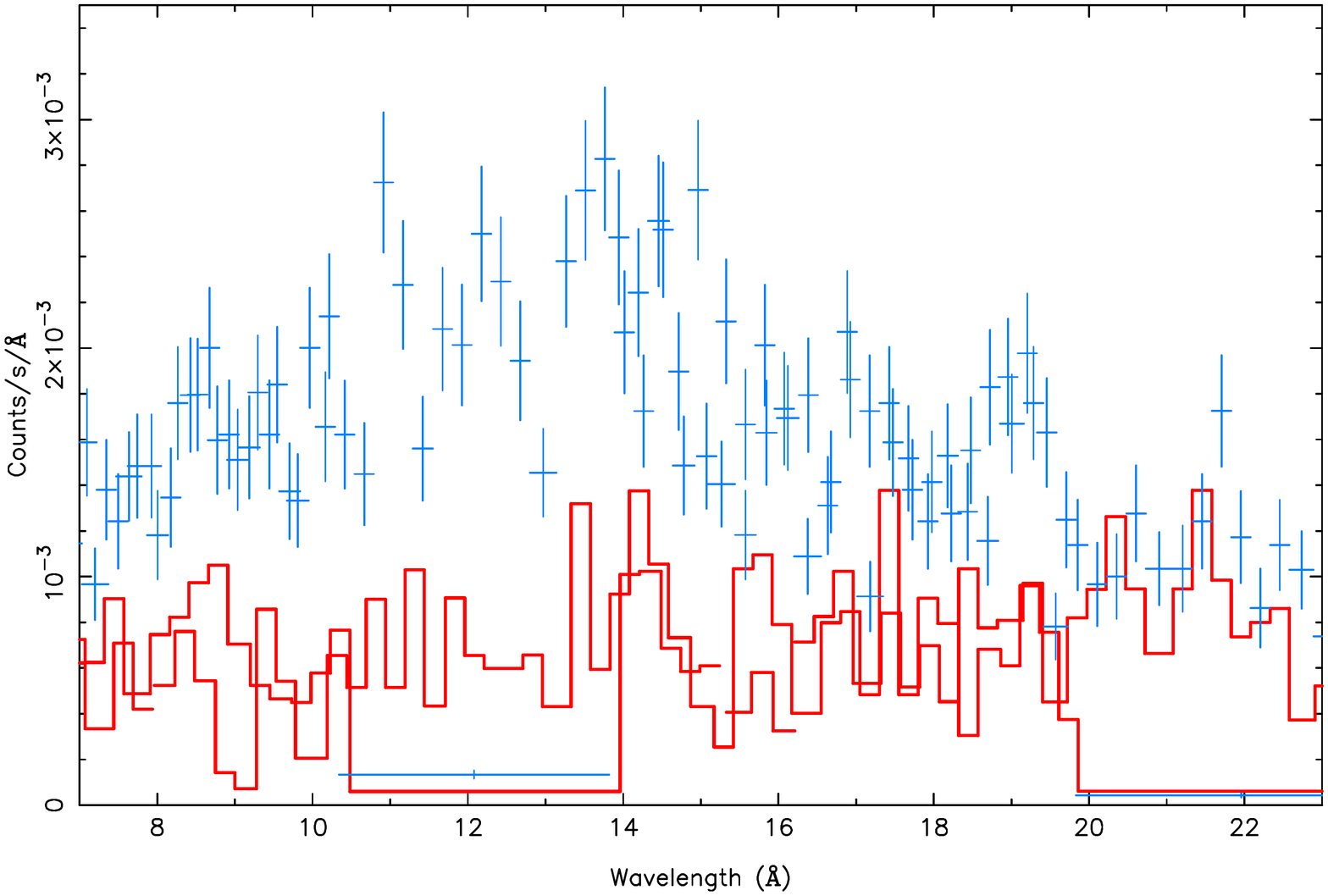}
    \caption{Blue - source (background not subtracted) spectrum of NGC 6946 X-1 between 7 and 23 \AA, red - background spectrum of the same object. Both RGS 1 and RGS 2 detector spectra are shown.}
    \label{N6946_bkg}
\end{figure*}

%%%%%%%%%%%%%%%%%%%%%%%%%%%%%%%%%%%%%%%%%%%%%%%%%%

% Don't change these lines
\bsp	% typesetting comment
\label{lastpage}
\end{document}